\documentclass[journal]{IEEEtran}
\usepackage{cite}

\ifCLASSINFOpdf
\else
\fi
\usepackage{amsmath}
\usepackage[nohyperlinks,printonlyused]{acronym}
\usepackage{url}
\usepackage{xurl}

\usepackage{graphicx}
\usepackage{amssymb}
\usepackage{multirow}
\usepackage{color}
\usepackage{tabularx}

\usepackage{array}
\usepackage{graphicx}
\usepackage{pifont}
\usepackage{comment}

\usepackage{etoolbox}

\patchcmd{\IEEEbiographynophoto}
  {\vskip 4\baselineskip plus 1fil minus 0\baselineskip}
  {\vskip 3pt\relax}
  {}
  {\PackageError{compact-bios}{Patch non applicata}
    {Verificare la definizione di IEEEbiographynophoto.}}

\newif\ifONDTlogiccolors
\ONDTlogiccolorstrue
\definecolor{ONDTcorecolor}{RGB}{24,74,135}
\definecolor{ONDTessentialcolor}{RGB}{27,112,70}
\definecolor{ONDTdetailcolor}{RGB}{156,103,0}

\begin{document}
\bstctlcite{BSTcontrol}

%

\title{Optical Network Digital Twin: From Digital Coherent Transmission to Intelligent Network Operations}





%
%
%

\author{Hideki~Nishizawa, Enrico~Miotto, Andrea~Rosso, Takahito~Tanimura, Zehao~Wang, Mingzhe~Han, Hiroyuki~Ishihara, Andrea~D'Amico, Agastya~Raj, Rocco~D'Ingillo, Takeo~Sasai, Giacomo~Borraccini, Toru~Mano, 
        Renato~Ambrosone, Behnam Shariati, Oscar~González-de-Dios, Arturo Mayoral López de Lerma, Jun-ichi~Kani, Manabu~Yoshimoto, Tingjun~Chen, Marco~Ruffini, Daniel Kilper, and~Vittorio~Curri
\thanks{H. Nishizawa, H. Ishihara, T. Mano, T. Sasai, J. Kani, and M. Yoshimoto are with NTT Laboratories, Japan.
E. Miotto, A. Rosso, R. D'ingillo, R. Ambrosone and V. Curri are with Politecnico di Torino, Italy.
T. Tanimura is with Hitachi, Japan.
Z. Wang, M. Han, T. Chen are with Duke University, USA.
A. D’Amico is with NEC Laboratories America, USA.
A. Raj, G. Borraccini, M. Ruffini, and D. Kilper are with Trinity College Dublin, Ireland.
B. Shariati is with Fraunhofer HHI, Germany.
O. Gonz´alez-de-Dios is with Telefonica Innovación Digital, Spain.
A. Mayoral López de Lerma is with Telecom Infra Project, Spain.
}}

\maketitle

\begin{abstract}
Optical Network Digital Twins (ONDTs) are emerging as a key framework for representing, synchronizing, and utilizing the physical state of optical networks for quality-of-transmission (QoT) estimation, optimization, and intelligent operation.
This survey provides a comprehensive view of ONDTs, spanning digital coherent transmission and physical-layer observability to intelligent network operations. We systematically organize ONDT technologies from ecosystem, functional, control, and implementation architecture perspectives and review the enabling technologies
required to construct, synchronize, and utilize digital representations of physical optical networks.
As a unifying framework to describe both ONDT design and validation, we introduce a categorization based on the following figures of merit: Fidelity, Dynamicity, and Deployability (FDD). In particular, we emphasize fit-for-purpose Fidelity, where modeling effort focuses on the physical effects of components rather than solely maximizing model accuracy, favoring scalability and generalizability.
Furthermore, experimental and field validations are examined to evaluate the practical applicability of ONDT technologies.
We further examine experimental and field validations, from physics-based QoT estimation and measurement-informed model refinement to field synchronization, service provisioning, multi-domain observability, and infrastructure characterization.
These validations highlight distinct deployment challenges: in conventional optical networks, ONDTs must demonstrate additional value while coexisting with established operational systems, whereas in emerging services and ecosystems, ONDTs can enable new services across heterogeneous administrative and resource boundaries.
Finally, we discuss the complementary roles of ONDTs and AI agents toward autonomous optical networks and identify key research directions for practical deployment.
\end{abstract}




\section{Introduction}
\label{Introduction}
Digital twin (DT) technology collects real-world data and enables simulation, prediction, optimization, fault location, predictive maintenance, condition monitoring, process control, and performance evaluation.
It is emerging as a means to manage the inherent complexity and uncertainty of large-scale physical systems.
Recently, the emergence of Physical Artificial Intelligence (AI) has increased the need for reliable digital representations of real-world systems. DT technology is increasingly recognized as a key enabler for achieving secure and reliable Physical AI systems.
In this perspective, it is essential to clearly articulate the value gained after DT adoption in terms of the initial investment costs required to develop and build DT.
To this end, one important requirement is Fidelity, referring to the ability of a DT to capture and faithfully reproduce the behavior of the target physical system.
Sufficient Fidelity for target applications enables reliable analysis, optimization, and operational decisions about physical systems.
%
This feature is particularly important in scenarios where potential impact of actions must be evaluated and explained beforehand, such as cases in which a potential impact can affect multiple users or spread between interconnected systems or in which system restoration would require a relatively long time. \begin{figure}[t!]
  \centering
  \includegraphics[width=1.0\hsize]{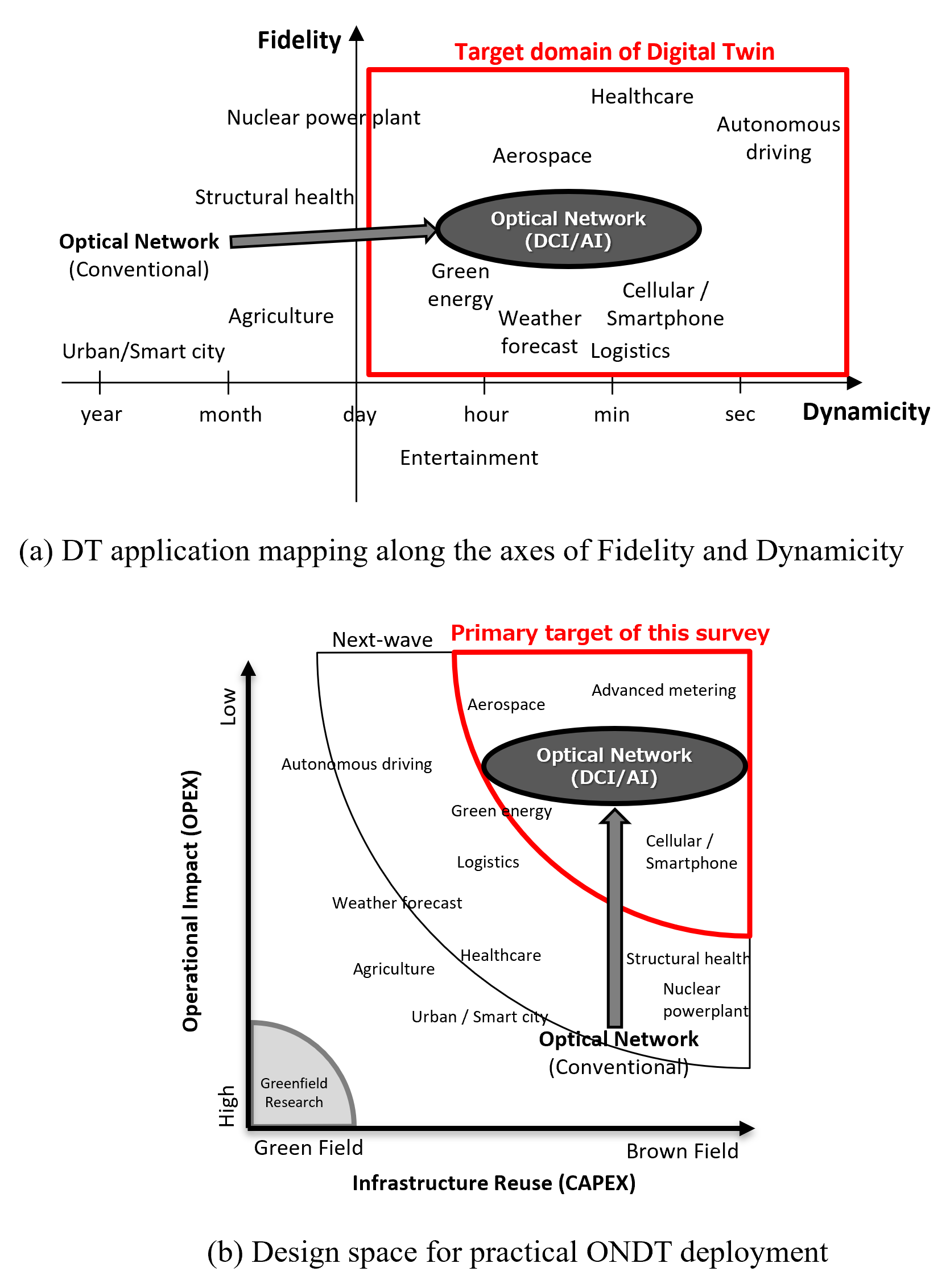}
  \caption{(a) DT application mapping and (b) design space for practical ONDT deployment.
  }
  \label{fig:dt_application_mapping}
\end{figure}
The second important requirement here is Dynamicity.
Dynamicity appropriate to target applications can provide greater operational benefits and a clearer return on investment.
While the advantages of Dynamicity depend on the application targeted by DT, high Dynamicity can open the way to new applications and even enable fundamentally new operational formats that could not be achieved without the benefits provided by a DT architecture.
Fig.~\ref{fig:dt_application_mapping}(a) shows DT application mapping along the axes of Fidelity and Dynamicity.
The emergence of new connectivity demands driven by datac enter interconnects (DCI) and AI infrastructures has created new opportunities for Optical Network Digital Twin (ONDT) adoption, especially in applications that require both high fidelity and fast synchronization.
The third important requirement is Deployability.
Fig.~\ref{fig:dt_application_mapping}(b) shows the design space for practical ONDT deployment.
The horizontal axis represents the extent of infrastructure reuse and associated capital expenditure (CAPEX), while the vertical axis represents operational impacts such as changes to operation/business support systems, operational workflows, and workforce training.
%
High deployability is achieved by maximizing infrastructure reuse while minimizing operational impact, as highlighted in Fig.~\ref{fig:dt_application_mapping}(b).
Especially, deploying ONDT to support emerging DCI/AI services over existing carrier optical line systems represents an important practical use case.
ONDT deployment does not inherently require the addition or replacement of physical network elements.
It is also important that ONDT can be easily applied to upgrading existing infrastructure, such as carrier line systems, because deploying new optical-fiber infrastructure requires not only equipment and materials but also large-scale civil engineering work and labor, resulting in substantial costs and deployment time.
%
Historically, carrier optical networks have evolved as the base layer of carrier-grade infrastructures supporting a huge number of users on Fiber to the Home (FTTH) and cellular services, so they have had to be highly reliable and robust.
Meanwhile, carriers must continuously extend, upgrade, and maintain the infrastructure while satisfying the Service Level Agreement (SLA) for an ever-increasing number of users.
%
%
\subsection{Motivations and Drivers for the Emergence of ONDT}
\label{Motivations and Drivers for the Emergence of ONDT}
Having outlined the requirements for a practical ONDT deployment, we now discuss the technological advances and emerging demands driving its adoption.
\subsubsection{Trends in optimization technology}
Coherent transponders can provide rich physical-layer telemetry and standardized management interfaces that can be leveraged by ONDT applications without requiring changes to the underlying line system (see Sec.~\ref{Standardization Toward Open and Programmable Optical Networks}). Leveraging their existing observability and programmability provides a practical entry point for ONDT-based operations while minimizing changes to the existing infrastructure.
Digital coherent transmission also simplifies optical network design by enabling receiver-side chromatic dispersion (CD) compensation using digital signal processing (DSP) and eliminating the need for dedicated optical dispersion compensation elements.
Gaussian Noise (GN) models reduce the computational complexity of estimating fiber nonlinear impairments and enable network-level design and optimization.
The digitalization of coherent transceivers (TRxs) has also accelerated the standardization of management interfaces and data models, enabling telemetry and software-defined network operations as described later in Sec.~\ref{Conceptual}.
\subsubsection{Evolution of Data Center Connectivity}
Conventional optical networks have traditionally been designed and deployed based on long-term traffic forecasts and capacity planning. Once deployed, their physical configuration typically remains relatively static for long periods of time and changes primarily due to increased traffic and network failures~\cite{soumplis2018}.
However, with the rapid growth of AI and cloud computing, the factors that determine connectivity demand are beginning to change. In distributed computing infrastructures for AI, machine-to-machine traffic is increasingly influenced by constraints such as regional power availability, where data must be stored and processed, and data sovereignty requirements, in addition to the placement and utilization of computing resources. These constraints are driving the geographical distribution of compute and data resources across hyperscale, regional, and on-premises datacenters.
Of particular relevance to coherent optical transport networks is the emergence of scale-across connectivity, which interconnects geographically distributed AI computing resources. Although large-scale AI training across multiple data centers has not yet become a widely deployed commercial use case, advanced demonstrations are already emerging~\cite{nvidia2025turbocharge}. Moreover, this evolution involves not only the spatial distribution of computing and data resources but also temporal variations in their utilization.
Measurements from production hyperscale cloud data centers have demonstrated clear daily periodicity, including traffic peaks and valleys over the course of a day, as well as application-dependent traffic distributions. Real traffic is also highly non-uniform and can exhibit hot spots, motivating monitoring-based bandwidth allocation and network reconfiguration \cite{Yan2023HyperscaleTraffic}.
%
Also, concerns about data security and privacy are driving governments and enterprises to adopt distributed sovereign cloud architectures \cite{Hoofnagle2019GDPR}.
Given this trend, the geographical distribution of DCs is accelerating, and as a result, the demand for large-capacity and flexible DCI networks is rapidly increasing \cite{Nishizawa2026JOCN}.

\subsubsection{Trends in Network Operation, “Common Metrics”}
Unlike packet networks, optical transport networks must account for analog physical-layer impairments that directly affect transmission performance. While standardized metrics such as throughput, delay, and packet loss can generally be used in a packet network, the quality of an optical transport system depends on physical-layer parameters such as Optical Signal-to-Noise Ratio (OSNR) and accumulated CD.
As optical networks move toward multi-vendor and disaggregated architectures~\cite{Lopez2020Programmable,TIP2021MUST}, differences in measurement methods across vendors and device generations demand vendor-independent performance assessments. Against this background, common physical layer metrics and vendor-independent quality of transmission (QoT) estimation techniques have emerged.
In particular, the generalized signal-to-noise ratio (GSNR) provides a common performance metric that integrates amplified spontaneous emission (ASE) noise and nonlinear interference (NLI), enabling consistent QoT estimation in heterogeneous network environments when combined with analytical models such as the GN model~\cite{Curri2022GNPyPhysicalLayer}.
The need for common metrics also promoted open and vendor-neutral design tools and software ecosystems, providing reproducible techniques for optical network design, testing, and operation.
As a result, common physical layer metrics, an open QoT model, and a vendor-neutral planning framework became key enablers for data-driven optical network operations and later provided a foundation for ONDT.
\subsection{Related Surveys}
\label{Related Surveys}
ONDT research has evolved from initial concepts to enabling technologies, laboratory demonstrations, and field trials, accompanied by a growing body of survey literature.
Table I classifies past surveys in terms of conceptual framework, enabling technologies, and validation to clarify the scope and contributions of this survey.
Wang \textit{et al.} \cite{Wang2021RoleDTOptical} propose an initial framework for applying the DT concept to optical communication systems focusing on fault management using deep learning techniques, hardware configurations, and transmission modeling.
As initial application-oriented research, this paper primarily focuses on specific technical functions and demonstrations of feasibility. Its scope complements later ONDT research, that is, studies on deployment requirements, software-defined operation frameworks, and large-scale field validations.
Zhuge \textit{et al.} \cite{Zhuge2023BuildingDTOptical} present one of the initial and comprehensive tutorials on DT for intelligent optical networks organizing this field about the three key technologies of modeling, telemetry, and self-learning.
As a technology-oriented tutorial, this research mainly focuses on basic constituent elements and technical abilities deemed necessary for implementing ONDT. The authors list accuracy, reliability, real-time synchronization, and future forecasting as key characteristics of ONDT. At the same time, they recognize the complexity of ONDT deployment, its feasibility in actual operation, and costs as key viewpoints in practical DT implementations and assigns them for future study.
Faruk \textit{et al.} \cite{Faruk2024MeasurementInformedDT} provide a comprehensive tutorial on physical-layer modeling and digital twins based on measurement information targeting optical fiber communication systems. This paper provides a detailed discussion of measurement-informed physical-layer modeling and outlines basic DT requirements, related network architecture, and operation applications.
Wang \textit{et al.} \cite{Wang2024DigitalTwinOpticalReview} provide a comprehensive review of recent advances in digital twin technologies for optical networks, covering architectural frameworks, physical-layer modeling, QoT estimation, monitoring mechanisms, and AI-assisted optimization techniques. While it offers a broad technical overview of digital twin components and their evolution, the focus remains primarily on technological aspects, without detailed discussion of deployment feasibility, economic considerations, or large-scale field validation in carrier-grade environments.
Zhuge \textit{et al.} \cite{Zhuge2025DigitalTwinAIAgent} present a tutorial that comprehensively covers DT and large language model (LLM)-based AI agents toward the implementation of autonomous optical networks. This tutorial, however, focuses on autonomous optical networks combining DT and AI agents.
Cruzes \textit{et al.} \cite{Cruzes2025OpticalAutomation} present a survey that targets optical network automation based on telemetry and takes up machine learning, digital twins, LLM, and agentic AI in an integrated manner. Its main focus, however, is on telemetry-based autonomous network operation and integration with agentic AI.
Nishizawa \textit{et al.} \cite{Nishizawa2025_ONDT, Nishizawa2026JOCN} discuss ONDT from the viewpoint of its practical deployment in optical transport networks and a system-level framework for integrating ONDT technology in the interconnection of distributed data centers using the Data Center Exchange (DCX) as a specific application scenario, respectively.

We can see from the above discussion that existing studies provide a systematic survey of key aspects making up ONDT including physical-layer modeling and monitoring, full-cycle ONDT technologies, autonomous operation using AI agents, telemetry-driven network automation, and practical deployment and field validation. On the other hand, each of these surveys focuses on specific technology domains, architectures, or deployment scenarios. None aim to establish links across ONDT conceptual development, architecture, physical-layer modeling, monitoring and synchronization, agentic AI, and practical deployment from the viewpoint of consistent system design.
In particular, these surveys do not sufficiently establish a framework that positions the interdependent requirements of FDD as common design axes in designing and evaluating a practical ONDT and that comprehensively analyzes the tradeoffs inherent to different technologies and architectures.

\begin{table}[htbp]
\caption{Comparison of Related Surveys on Optical Network Digital Twin}
\label{tab:related_surveys}
\centering
\scriptsize
\setlength{\tabcolsep}{3pt}
\renewcommand{\arraystretch}{0.9}

\begin{tabular*}{\columnwidth}
  {@{\extracolsep{\fill}}|c|cc|cccc|cc|}
\hline
\multirow{3}{*}{\textbf{Ref.}}
& \multicolumn{8}{c|}{\textbf{Contributions}} \\ \cline{2-9}

& \multicolumn{2}{c|}{\textbf{Conceptual}}
& \multicolumn{4}{c|}{\textbf{Enabling Tech.}}
& \multicolumn{2}{c|}{\textbf{Validation}} \\ \cline{2-9}

& Hist.
& Arch.
& Model.
& Mon.
& Ctrl.
& AI
& Deploy.
& Field \\
\hline

{\cite{Wang2021RoleDTOptical}}
& $\times$
& $\checkmark$
& $\checkmark\checkmark$
& $\checkmark$
& $\checkmark$
& $\times$
& $\times$
& $\times$ \\
\hline

{\cite{Zhuge2023BuildingDTOptical}}
& $\times$
& $\checkmark$
& $\checkmark\checkmark$
& $\checkmark\checkmark$
& $\checkmark\checkmark$
& $\times$
& $\times$
& $\times$ \\
\hline

{\cite{Faruk2024MeasurementInformedDT}}
& $\checkmark$
& $\checkmark$
& $\checkmark\checkmark$
& $\checkmark\checkmark$
& $\checkmark$
& $\times$
& $\checkmark$
& $\checkmark$ \\
\hline

{\cite{Wang2024DigitalTwinOpticalReview}}
& $\checkmark$
& $\checkmark\checkmark$
& $\checkmark\checkmark$
& $\checkmark\checkmark$
& $\checkmark\checkmark$
& $\checkmark$
& $\times$
& $\checkmark$ \\
\hline

{\cite{Zhuge2025DigitalTwinAIAgent}}
& $\checkmark$
& $\checkmark$
& $\checkmark\checkmark$
& $\checkmark\checkmark$
& $\checkmark\checkmark$
& $\checkmark\checkmark$
& $\times$
& $\checkmark\checkmark$ \\
\hline

{\cite{Cruzes2025OpticalAutomation}}
& $\times$
& $\checkmark$
& $\checkmark$
& $\checkmark\checkmark$
& $\checkmark\checkmark$
& $\checkmark\checkmark$
& $\checkmark$
& $\times$ \\
\hline

{\cite{Nishizawa2025_ONDT}}
& $\times$
& $\checkmark$
& $\checkmark$
& $\checkmark$
& $\checkmark$
& $\times$
& $\checkmark\checkmark$
& $\checkmark$ \\
\hline

{\cite{Nishizawa2026JOCN}}
& $\times$
& $\checkmark$
& $\checkmark$
& $\checkmark\checkmark$
& $\checkmark\checkmark$
& $\times$
& $\checkmark$
& $\checkmark\checkmark$ \\
\hline

This survey
& $\checkmark\checkmark$
& $\checkmark\checkmark$
& $\checkmark\checkmark$
& $\checkmark$
& $\checkmark$
& $\checkmark$
& $\checkmark\checkmark$
& $\checkmark\checkmark$ \\
\hline
\end{tabular*}

\vspace{2pt}
\parbox{\columnwidth}{%
\raggedright
\scriptsize
\textit{Abbreviations:}
Hist. = History;
Arch. = Architecture;
Model. = Modeling;
Mon. = Monitoring and Synchronization;
Ctrl. = Control and Actuation;
AI = Agentic AI;
Deploy. = Deployability;
Field = Field Validation.

\textit{Symbols:}
$\checkmark\checkmark$: systematic and broad coverage of the subject;
$\checkmark$: substantial but focused or partial coverage;
$\times$: subject not addressed or only briefly mentioned.
}
\end{table}

\subsection{Survey Contributions}
\label{Survey Contributions}
To fill this gap, we organize this survey systematically from the historical development of ONDT to four architectural design dimensions, physical-layer modeling, monitoring and synchronization, agentic AI. We also analyze the mutual relationships and tradeoffs among the requirements of FDD using them as a common design framework, and establish a unified ONDT paradigm that bridges the gap between physical layer digital modeling and AI agent-assisted network operations.
This survey examines how technological advances collectively enable ONDT and how their integration supports intelligent optical network operations. Its main contributions are summarized in four points.

\textbf{Evolutionary view of optical network observability} We reinterpret the historical evolution of optical networking as a process of incrementally improving network observability through digital coherent transmission, analytical modeling, open standardization, telemetry and digital longitudinal monitoring (DLM), data-driven technologies, and Agentic AI. We reveal how these technologies comprehensively formed the technical basis for practical ONDT.

\textbf{Systematization of multifaceted ONDT architecture} We organize the ONDT architecture into four complementary architectural design dimensions called Ecosystem architecture, Functional architecture, Control architecture, and Implementation architecture. These dimensions provide a common design space for different ONDT approaches.

\textbf{Unified design framework based on FDD} We introduce FDD as three fundamental design requirements for practical ONDT.
These are Fidelity for purpose-dependent physical layer accuracy, Dynamicity for application-dependent synchronization, and Deployability for incremental integration into existing network environments.
This framework provides a common basis for analyzing technical and architectural trade-offs and evaluating practical deployment strategies across optical network scenarios.

\textbf{Bridge the gap between physical phenomena and AI-agent-assisted network operation} We systematically link physical layer degradation factors, digital models, monitoring/telemetry, and AI agent-assisted network operations to reveal how complementary technologies enable reliable ONDT. We distinguish the historical evolution of AI from the actual design of AI agent-assisted ONDT and systematically organize the latter into components including agent architecture, tools, safety, and operational workflows.

\subsection{Survey Structure}
\label{Survey Structure}
This paper is organized as follows.
Section \ref{Conceptual} introduces basic concepts of ONDT and its problem space and examines the historical evolution of optical network technologies from the viewpoint of improved visibility.
This section also organizes ONDT architecture into the four complementary design dimensions, and clarifies the relationship between the physical network and its digital representation and network operation.
Continuing on, this section defines FDD as basic design requirements for achieving a practical ONDT and presents a common design framework for analyzing future technologies and architectures.
Section \ref{Enabling Technologies} reviews main enabling technologies and analyzes the features, scope, and mutual relationships of each technology from the viewpoint of FDD.
The latter part of the section \ref{Enabling Technologies} focuses on AI-agent-assisted autonomous optical network operation and systematically discusses agent architecture, inference/planning, tool use, collaboration with network operation, safety, and practical deployment considerations.
Section \ref{Validation} discusses the aspects of validation and practical deployment including use cases and field demonstrations and focuses on filling in the gap between technical approaches and real-world optical network operation/industry.


\section{Conceptual}
\label{Conceptual}
\subsection{Evolution of Optical Network Digitalization}
\label{Evolution}
Progress in optical network digitalization has not simply improved transmission performance in terms of longer distances and higher capacities.
It has also fundamentally changed the way optical networks are designed and optimized, while improving \emph{network observability} in a stepwise manner.
%
Combined with subsequent standardization activities owing to the participation of a variety of stakeholders, this improvement in observability accelerated the development of analytical QoT models, vendor-independent performance metrics, standardized telemetry interfaces, and open software frameworks.

\subsubsection{From Digital Coherent Reception to GN Modeling}
\label{From Digital Coherent Reception to GN Modeling}
Prior to 2010, optical transmission systems were mainly based on analog technology and the intensity-modulation/direct-detection (IM-DD) scheme at 10 Gbps per wavelength.
In the transition to 100 Gbps, IM-DD has physical-layer limitations such as CD, polarization mode dispersion (PMD), and ASE noise, and it was not realistic to realize 100 Gbps transmission by simply expanding the conventional IM-DD architecture.
The introduction of digital coherent transmission \cite{Kikuchi2006Coherent} using dual-polarization quadrature phase-shift keying (DP-QPSK) marked a paradigm shift in the industry.
Advanced DSP algorithms that include CD compensation and adaptive equalization in the frequency domain can dynamically compensate for CD and PMD.
Phase noise, CD, polarization mode dispersion, nonlinear distortion, TRx degradation, and filtering effects can be represented by digital samples and processed by adaptive DSP algorithms~\cite{Savory2008DigitalFilters}. 


\begin{figure*}[ht!]
  \centering
  \includegraphics[width=1.0\linewidth]{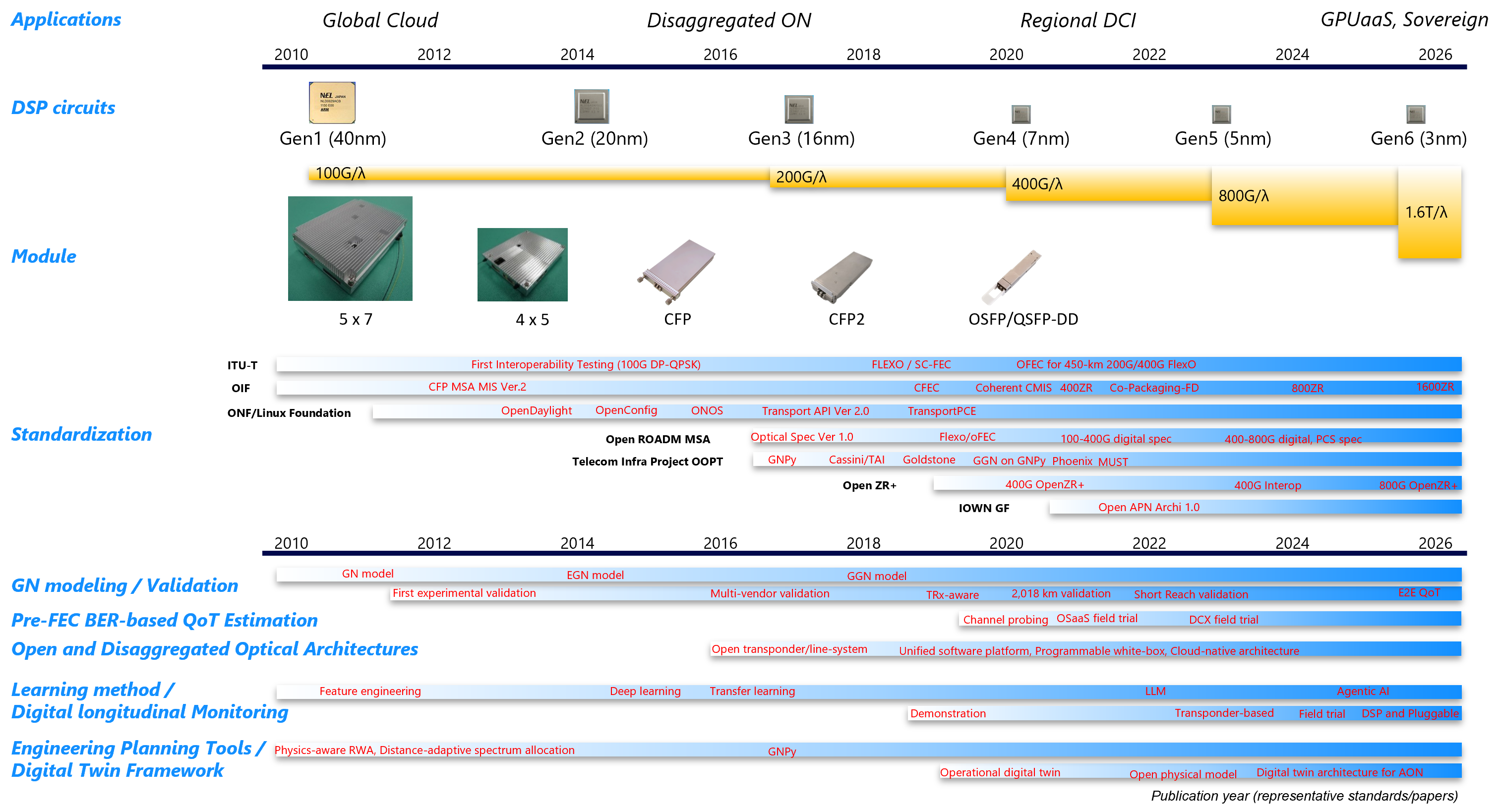}
  \caption{Historical development of key technologies contributing to ONDT.}
  \label{fig:evolution_history}
\end{figure*}
%
Digital coherent transmission has also enabled a substantially simplified approach to modeling fiber propagation.
For a generic field propagating in a single-mode fiber, the reference model is the coupled nonlinear Schrödinger equation (CNLSE) \cite{AGRAWAL2013}.
This model assumes that the fundamental fiber mode is perturbed only by the Kerr nonlinear effect, an assumption well satisfied in telecommunication fiber usage, and includes the birefringent nature of real fibers and its stochastic behavior.
Therefore, in its most general form, the CNLSE is a stochastic differential equation that is extremely difficult to handle analytically.
By using DP signals to carry data, the physical paradigm changes substantially.
Also thanks to the high symbol rate, each wavelength division multiplexing (WDM) channel can be considered fully depolarized over distances shorter than the characteristic length of the birefringence random process.
Consequently, the overall propagating modal amplitude can be regarded as depolarized, allowing birefringence to be averaged out.
Moreover, thanks to DSP, PMD and polarization rotations are compensated for at the receiver \cite{savory2010jstqe}.
These features enable the use of the Manakov equation (ME) \cite{Marcuse1997} to model fiber propagation.
This represents a substantial simplification, since the ME is a deterministic differential equation including loss, CD, and the Kerr effect.
The ME is formally a local approximation in the frequency domain, but for DP signals its use can be extended to wideband scenarios \cite{Cantono:18}.
In this case, the frequency dependence of loss and nonlinear coefficient must be included, together with stimulated Raman scattering (SRS) \cite{rizzi2026}.

With the introduction of DP coherent technologies in the late 2000s and early 2010s, the research community started to investigate their specific implications for fiber propagation.
In \cite{curri2008ptl}, it was shown that in-line CD compensation is inconvenient for DP coherent technologies.
This clarified that the performance assessment of these technologies is much simpler than in IM-DD scenarios: fiber characteristics become clear system figures, whereas IM-DD systems required the optimization of dispersion maps.
In 2010, Carena \textit{et al.} showed for the first time the Gaussian nature of the interference generated by fiber nonlinearity, later named NLI, together with its practical interpretation as additive noise at the receiver~\cite{carena2010ecoc}.
This result proposed the model of a transparent lightpath as an additive GN nonlinear channel, also considering that ASE noise from optical amplifiers is well modeled as additive GN~\cite{essiambre2010jlt}.
Contrary to IM-DD scenarios, this analysis clearly displayed an algorithmic dependence of performance on system figures such as fiber characteristics, symbol rate, channel spacing, and occupied bandwidth.
These developments laid the foundation for the GN model, which enables system-performance estimation with practical computational complexity, proposed by Poggiolini \textit{et al.} in 2011~\cite{poggiolini2011ptl}.
Within this framework, an important distinction concerns the accumulation of NLI over multiple fiber spans.
The coherent GN model sums nonlinear fields across spans while preserving their relative phases, retaining inter-span coherence.
The incoherent GN (IGN) approximation instead sums per-span NLI powers, neglecting inter-span coherence. Generally accurate for many practical dispersion-uncompensated links, it enables span-by-span disaggregation and simpler accumulation rules, making it particularly suitable for network-level QoT estimation~\cite{Poggiolini_GN_Derivation,Carena2012ModelingNLI}.
Although formally distinct, the IGN approximation is commonly referred to as the GN model in network-level QoT estimation. For clarity, we adopt this naming convention throughout the paper, unless coherent accumulation is explicitly considered.

%
The strength of the GN model is also its weakness: it describes NLI as four-wave mixing among spectral components while neglecting the digital nature of each WDM channel, the modulation-format-dependent signal constellation, and the resulting span-by-span statistical correlations.
At the same time, independent analyses of NLI showed that the ideal Gaussian-noise approximation is not exact: NLI can exhibit modulation-format dependence that explains discrepancies between GN predictions and detailed simulations~\cite{Dar:13,Dar:14}.
To overcome these limitations, the enhanced GN (EGN) model, was proposed \cite{Carena:14}.
The EGN model corrects the GN model by adding correction terms that accounts for the Gaussianization process, modulation-format dependence, and correlation effects.

Besides enabling a quick assessment of the amount of NLI introduced by fiber propagation, the GN model makes it possible to easily derive the scaling laws of NLI with respect to system parameters.
The standard GN model relies on constant fiber parameters; therefore, it does not include the frequency dependence of loss and nonlinear coefficient, and it neglects SRS.
To extend the use of the GN model beyond the C band and to enhance its accuracy also within the C band, the generalized Gaussian-noise (GGN) model was proposed by Cantono \textit{et al.} in 2018 \cite{Cantono2018GGN}, and later optimized and refined by D'Amico \textit{et al.} in 2022 \cite{damicoJLT2022}.
In parallel with the derivation of the GGN model, Semrau \textit{et al.} proposed the inter-channel stimulated Raman scattering Gaussian-noise (ISRS-GN) model in \cite{Semrau:18}, later refined in \cite{semrau2020modeling}.
 
Although GN-model-based QoT estimation was initially developed primarily to evaluate line-system impairments, ASE noise and NLI noise, estimating the end-to-end (E2E) quality of an actual optical lightpath in terms of BER, Q-factor, or margin also requires TRx implementation impairments to be considered.
In the early 2010s, TRx effects were therefore incorporated into QoT estimation frameworks as fixed implementation penalties or equivalent additive Gaussian noise~\cite{Torrengo:11, vacondio_2012_nonlinear, chin_2017_probabilistic}, enabling ASE, NLI, and TRx noise to be combined within a common SNR or GSNR representation~\cite{berg_2019_demystifying}.

These developments established the foundation for an efficient physical-aware QoT estimation, allowing network design and lightpath feasibility evaluation with a limited set of parameters, a key feature for ONDT applications.
\subsubsection{Standardization Toward Open and Programmable Optical Networks}
\label{Standardization Toward Open and Programmable Optical Networks}
Digital coherent reception not only improved transmission performance but also fundamentally changed optical transmission to a “digitally observable system \cite{Tanimura2016DNN, Tanimura2019CNN}.” By converting measurable degradation factors in optical transmission to digital quantities by coherent DSP, it has become possible to continuously observe performance metrics in relation to CD, polarization effects, and forward error correction (FEC).
This became the first step toward Software Defined Networking (SDN) operation in optical transmission systems and later established a technical foundation for the provision of a common framework in relation to vendor-independent performance through standardized FEC.
Through stepwise standardization as shown in Fig.~\ref{fig:evolution_history}, FEC has become a key enabler of multi-vendor interoperability.
In 2018, the International Telecommunication Union Telecommunication Standardization Sector (ITU-T) established a common foundation for interoperable coherent transmission by standardizing an OTU4 long-reach interface based on Stair-case FEC (SC-FEC) (Recommendation G.709.2)~\cite{ITU_G7092_2018} and flexible optical transport network (OTN) long-reach interfaces Recommendation G.709.3~\cite{ITU_G7093_2018} in a mutually complimentary manner.
%
The Optical Internetworking Forum (OIF) adopted Concatenated FEC (CFEC) as part of its 400ZR Implementation Agreement to achieve interoperable coherent pluggable in DCI applications against the background of requests from hyperscale cloud providers like Google and Microsoft~\cite{OIF_400ZR_IA_2022}.
Next, to support carrier transport requirements over longer distances, Open ROADM standardized open FEC (oFEC) for 200G/400G coherent transport~\cite{OpenROADM_WPort_Digital_v6_2023}.
After this, OpenZR+ Multi-Source Agreement (MSA)~\cite{OpenZRplusWhitePaper} built a bridge between the 400ZR and Open ROADM ecosystems by combining 400ZR Ethernet-oriented architecture and the high-gain oFEC defined by Open ROADM.
Another important contribution brought about by FEC standardization is the establishment of common physical-layer semantics between multi-vendor coherent systems.
Once interoperable FEC schemes were standardized, pre-FEC bit error rate (BER) changed from an implementation-specific parameter to a common QoT metric that could be consistently interpreted between different TRx implementations.
%

Continued advances in CMOS technology have enabled coherent DSP chips to achieve high integration and increased energy efficiency, allowing coherent TRx to evolve from large embedded modules to small pluggable form factors, as shown in the center of Fig.~\ref{fig:evolution_history}.
The evolution of coherent pluggable technology has expanded applications from traditional transport systems to routers and switches, and accelerated the adoption of internet protocol (IP) over dense wavelength division multiplexing (IPoWDM) architectures.
In parallel, coherent TRx has evolved from vertically integrated products to disaggregated architectures, enabling independent development of hardware and software~\cite{Nishizawa2026JOCN}.
In 2019, the Cassini whitebox switchponder demonstrated multi-vendor coherent TRx interoperability using open hardware abstraction interfaces, following the initial specification of the Transponder Abstraction Interface (TAI) in 2017.
This evolution has also facilitated the emergence of white-box optical transmission platforms that integrate cloud-native data center infrastructure and optical transport technologies~\cite{Nishizawa2021Whitebox, Lopez2020Programmable}.
%
%
%
%
\begin{figure}[hb!]
  \centering
  \includegraphics[width=1.0\hsize]{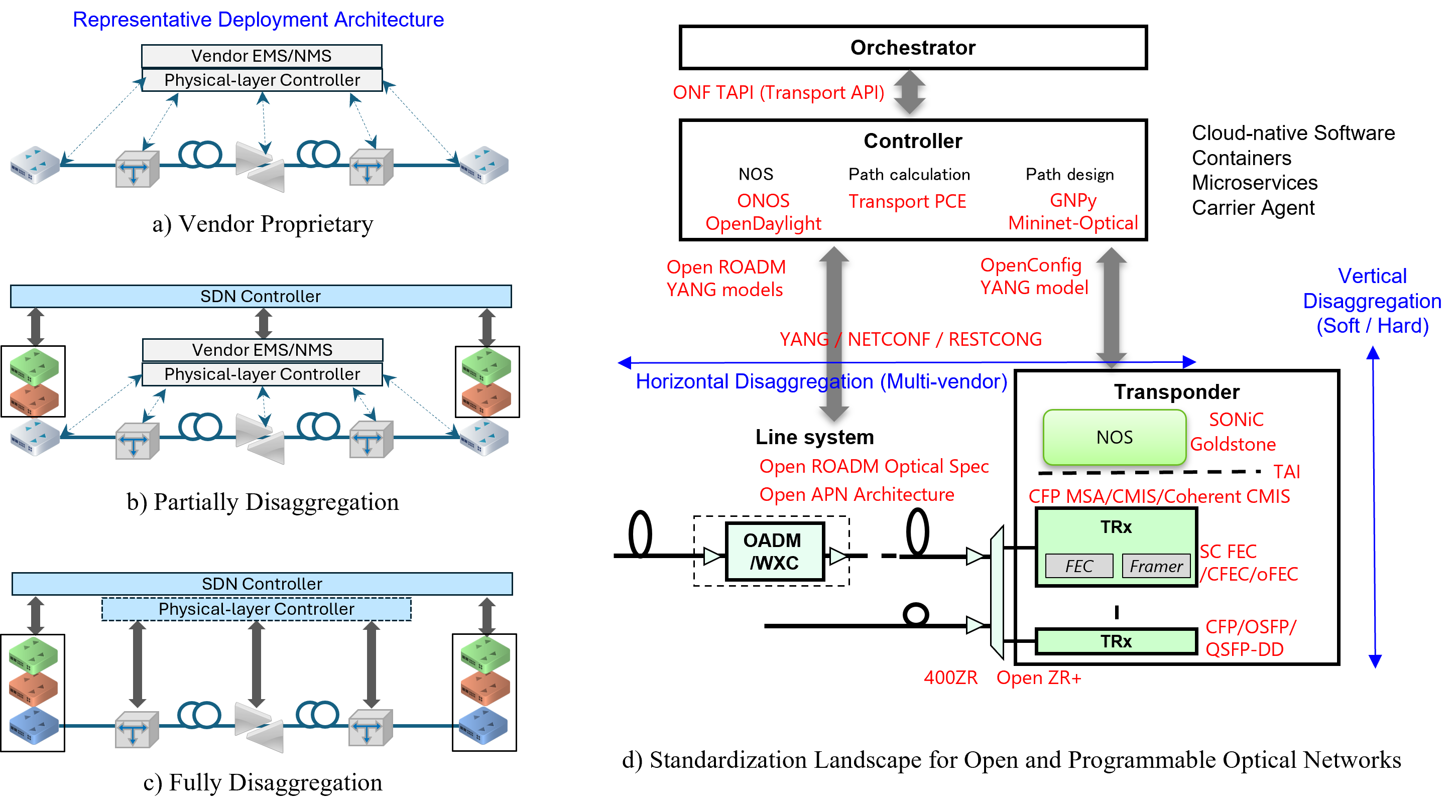}
  \caption{The evolution toward open and programmable optical networks along three complementary perspectives: representative deployment architectures, horizontal disaggregation across network elements and vendors, and vertical disaggregation between device hardware and software layers.}
  \label{fig:Open and Programmable Optical Network Architecture}
\end{figure}

Understanding the state of the network and predicting future transmission performance cannot be achieved simply on the basis of physical-layer observability.
To effectively use telemetry obtained from DSP in network design and operation, there is a need for a physical layer model that can map observable quantities to fiber propagation phenomena and efficiently estimate QoT under diverse transmission conditions.
%
Optical transport networks progressed from vertically integrated systems to more open, programmable architectures.
As operators came to deploy equipment from multiple vendors, a vendor-independent mechanism became indispensable for performance evaluation, configuration management, telemetry collection, and network control.
These demands prompted the development of common physical layer metrics as well as standardized interfaces, data models, and management frameworks across TRx, circuit systems, controllers, and orchestration platforms.
Fig.~\ref{fig:Open and Programmable Optical Network Architecture} illustrates the evolution to open and programmable optical networks through three representative deployment architectures and two complementary dimensions of disaggregation: horizontal and vertical~\cite{Lopez2020Programmable,TIP2021MUST}.
Horizontal disaggregation enables interoperability among transponders, optical circuit systems (OLSs), controllers, and orchestration platforms from multiple vendors through standardized interfaces and data models.
Vertical disaggregation, on the other hand, separates transponder hardware, OS, and software functions through hardware abstraction interfaces and management application programming interfaces (APIs).
This progression to open and programmable optical networks did not begin solely because of horizontal network disaggregation.
Starting with the standardization of digital coherent transmission, the development of interoperable FEC schemes, common interfaces for coherent DSP, and TRx interoperability toward multi-vendor optical transport systems became a key trigger for achieving vertical disaggregation.
In the following, we trace the development of interoperability and programmability across the two complementary disaggregation dimensions.
The introduction of management interfaces for coherent pluggable modules including CFP MSA- and common management interface specification (CMIS) based interfaces enabled the configuration and monitoring of coherent TRxs by standardized methods.
Along with the spread of SDN, the scope of standardization expanded from interoperability on the physical layer to programmable network control.
These developments enabled optical devices and transport networks to be integrated into software-defined operational frameworks.
Open ROADM, launched in 2016, focused on multi-vendor interoperability in optical transport systems through common specifications and data models, while Telecom Infra Project Open Optical and Packet Transport (TIP OOPT), also launched in 2016, promoted disaggregation architecture, open optical systems, and open-source software including open-source QoT estimation tool GNPy.
The rapid expansion of coherent pluggable optical technology described in Section~\ref{Standardization Toward Open and Programmable Optical Networks} further accelerated these activities. The deployment scope of interoperable coherent TRx has greatly expanded, strengthening the vendor-independent optical networking ecosystem.
While fully disaggregated optical networks have attracted significant research interest, partially disaggregated architectures are a practical deployment approach in terrestrial transportation networks.
One motivation is the relatively rapid evolution and substantial cost contribution of coherent transponders, as illustrated in Fig.~\ref{fig:evolution_history}.
Thus, introducing interoperable transponders while retaining the existing line system can improve Deployability and reduce transition costs, although the achievable openness and programmability remain constrained by line-system interfaces and operational access.
The Innovative Optical and Wireless Network (IOWN) Global Forum, established in 2020, further expanded the scope of open networking initiatives to focus on E2E optical connectivity, programmable network control, and digitized network operations. These activities further emphasized the importance of integrating physical network information with software-defined operational frameworks.
Overall, these standardization activities illustrate the stepwise evolution of optical networking from interoperable digital coherent transmission toward open and programmable operational frameworks.
%
\subsubsection{Why Optical Networks Remain Challenging for Digital Twin}
\label{Why Optical Networks Remain Challenging for Digital Twin}
Digital coherent transmission, GN-based QoT estimation, and open telemetry have greatly improved the observability and controllability of the optical network.
Nevertheless, applying digital twin technology to optical networks poses challenges distinct from those in packet networks, particularly in predicting E2E physical-layer transmission quality.
The reason for this is that E2E QoT is determined by the cumulative interaction of multiple physical-layer degradation factors distributed along fibers, amplifiers, reconfigurable optical add-drop multiplexers (ROADMs), and TRxs.
%
%
These physical phenomena can be broadly divided into two types.
The first type includes degradation factors like ASE, NLI, and receiver limitations whose physical origins can be relatively well modeled.
In many cases, the effects of these phenomena can be characterized by an analytical model, which means that they are applicable to design and optimization by software.
%
%
The second type of physical phenomena includes degradation factors caused by complex interactions in optical components, such as wavelength selective switch (WSS) filtering effects, input power dependency of the amplifier gain spectrum, inter-spectrum interaction within the fiber due to the Raman effect, and inter-channel power fluctuations related to polarization.
These effects generally have strong wavelength dependency and couple with other degradation factors.
In addition, some physical parameters required for analytical modeling may be unavailable or vendor-proprietary, making accurate analytical modeling difficult~\cite{Ferrari2020GNPy}.
%
%
Such OLS optimization technology is vendor proprietary, and its limited openness has become a barrier to interconnecting multiple domains with equipment from different vendors.
Recently, moreover, when accommodating multi-generation and multi-vendor TRxs each having different receive functions in a single line system, operation designed not on the basis of optimizing QoT for each link but rather on an E2E target QoT has been proposed~\cite{Nishizawa2025_ONDT}.
%
Distinguishing between these two types of physical phenomena is particularly important from the viewpoint of ONDT.
In the case that transmission performance can be mainly determined by degradation factors that are analytically easy to handle, sufficiently accurate predictions can be made according to QoT estimations based on physical models enabling practical software-driven optimization.
However, as interdependent and wavelength-dependent effects grow, maintaining sufficient Fidelity may require updated telemetry and model calibration, potentially complemented by data-driven approaches.
This issue is particularly pronounced in large-scale, high-load, and dynamically reconfigurable optical networks.
These degradation factors depend on the target application and on different physical media and network elements such as transmission fiber, optical amplifiers, WSS, and coherent TRxs, so ONDT will require different architectures and modeling techniques for each.
\subsubsection{A New Monitoring Paradigm: The TRx as a Distributed Sensor}
\label{A New Monitoring Paradigm: The Transceiver as a Distributed Sensor}
As physical-layer modeling and telemetry evolved, monitoring impairments distributed across fibers, amplifiers, and ROADMs became increasingly important for understanding the actual state of optical transmission lines. Monitoring approaches for capturing these impairments can be broadly classified into two categories.
The first deploys dedicated instruments at intermediate (relay) sites, representative examples being the optical channel monitor (OCM) and the optical time-domain reflectometer (OTDR). Dedicated monitors are rarely present at every intermediate site, and OTDR in particular is inherently confined to a single span, since its probe light cannot traverse the optical isolators within amplifiers and ROADMs; monitoring an entire link therefore requires a separate instrument at each span. Under in-service conditions, the probe light may also interfere nonlinearly with the data-carrying signal. In multi-operator (multi-domain) environments, moreover, intermediate sites beyond an administrative boundary are altogether inaccessible, so that no monitoring information can be obtained from them \cite{Kaeval2025OSaaS}.
With the widespread deployment of digital coherent transmission in the 2010s, the second category exploits the coherent TRx located at the link endpoint. Here, transmission-line parameters are estimated inversely from the impairment-compensation blocks within the receiver DSP, providing a low-cost means---requiring little additional hardware---of obtaining a variety of physical parameters such as CD, differential group delay (DGD), polarization-dependent loss (PDL), second-order polarization mode dispersion (SOPMD), and signal-to-noise ratio (SNR) \cite{Hauske2009OPM,Soriano2011CDest,Li2018PDLmonitor,Do2013SOPMD,Tanimura2022book}. Conventionally, however, these were obtained only as values accumulated over the entire link length, without the spatial resolution required to indicate where along the path they arose.
In 2019, Tanimura \textit{et al.} demonstrated a new monitoring approach that estimates the longitudinal optical-power profile of an entire multi-span link from signals acquired by a coherent receiver at the link endpoint, enabling distance-resolved localization of power anomalies without dedicated monitoring instruments along the transmission line~\cite{Tanimura2020AnomalyID, Tanimura2019PDP}. Later, Sasai \textit{et al.} extended this concept to a broader range of physical properties, including longitudinal fiber loss and CD profiles, amplifier gain spectra, and optical-filter responses, and these DSP-based techniques for revealing physical characteristics along the fiber propagation direction were collectively referred to as DLM~\cite{Sasai2022DLM}. The TRx functions not only as an endpoint to send and receive data, but also as a distributed sensor to estimate the internal physical state of the terminating link.
\subsubsection{From Learning Models to Agentic AI}
\label{From Learning Models to Agentic AI}
\paragraph{From Feature Engineering to Deep Learning}
\label{feature_engineering_to_deep_learning}
The evolution of AI in optical fiber communications and optical networks can be understood not just as a set of larger models, but as an incremental extension of what optical networks can observe, represent, infer, act, and ultimately predict~ \cite{Tanimura2022book}.
Before the widespread deployment of digital coherent transmission and rich telemetry infrastructure, AI applications in optical communication systems were largely constrained by the amount of information that could be measured or economically extracted.
The dominant paradigm in this era was feature engineering. Human experts designed compact descriptors of optical signals or network states, and machine learning models were trained on top of these descriptors. Classical machine learning methods were used for tasks including OSNR estimation, modulation format identification, fault classification, fault localization, and quality of transmission (QoT) prediction~\cite{Chan2010book}. Models trained on one transmission condition, modulation format, fiber type, or network configuration often failed to generalize to another. In other words, the bottleneck was not only the learning algorithm itself, but the fact that the system was being observed through a narrow, human-defined aperture.
The limitations of handcrafted feature engineering therefore motivated a transition toward representation learning, where informative features could be learned automatically from increasingly rich waveform- and network-level observations.
This transition was enabled by the advent of digital coherent receivers, radically expanding the observability of optical communication systems. This observability breakthrough created the conditions under which deep learning could become useful in optical communications~\cite{Tanimura2016DNN}. Deep learning shifted the role of AI from classification over handcrafted features to representation learning from raw or partially processed data \cite{Tanimura2022book}.
\paragraph{Foundation model and Model Adaptation}
\label{transfer_learning_model_adaptation}
Later, transfer learning and domain adaptation enabled the knowledge learned from one device, link, or network environment to adapt to different operating conditions with less training data, improving the generalization and deployability of AI models. Models trained on laboratory datasets or simulations may not generalize to field-deployed systems. A major challenge in deploying AI models in optical networks is the difficulty of collecting sufficient measurement data to retrain or fine-tune deep learning models in production environments, where large data acquisitions may disrupt normal network operations.
To address this limitation, analytical model adaptation methods have been proposed, where a small number of field measurements are used to refine model parameters and adapt analytical models to field-deployed optical systems \cite{Borraccini2025differential,yang2025input}. Similarly, transfer learning and cascaded learning approaches have been developed to adapt deep learning models for accurate E2E optical link power prediction using limited field measurements \cite{raj2024multi,Wang2025cascaded}. As the volume and variety of observed data continue to expand, training robust models requires not only systematically ensuring the quality of individual data points, but also a data pipeline that integrates and manages diverse data obtained at different timescales and granularities.
\paragraph{From LLMs to Agentic AI}
\label{ai_agent_optical_system}
As richer and more heterogeneous network information became available, foundation models enabled integration of operational data and domain knowledge, and inference models extended AI beyond statistical prediction to multi-stage inference and planning.
While deep-learning-based digital twins have greatly improved our ability to model, predict, and optimize optical network performance, they still require human operators or external controllers to translate their outputs into network actions.
Additionally, neural models can behave as black boxes, making it difficult for operators to trust them for safety-critical network control. However, these methods were generally designed for individual tasks and lacked the ability to autonomously coordinate multiple network operations.
The advent of generative artificial intelligence (GenAI), and LLM in particular, marked the next step by enabling natural language interactions with networked systems. One of the earliest approaches involved LLM directly manipulating optical networks~\cite{Tanimura2024LLM}. Nevertheless, early LLM-based approaches focused primarily on translating user requests into configuration commands or executable code and providing more intuitive interfaces for network management, but still relied on external control frameworks for execution~\cite{sun2025experimental}. More recently, AI agents have extended the role of LLM from language understanding to autonomous decision-making by integrating reasoning, task planning, tool invocation, and feedback-driven execution~\cite{wang2026agentic} (see Table~\ref{tab:fdd_comparison}). Instead of simply generating configuration scripts, AI agents can decompose high-level objectives into sequential actions, interact with heterogeneous network devices through standardized interfaces, monitor execution results, and iteratively refine decisions based on system feedback. This shift from predictive AI models and passive language models toward active agentic systems represents an important milestone toward self-managing optical networks capable of autonomous provisioning, monitoring, optimization, and fault recovery with minimal human intervention.
\subsubsection{Convergence toward Optical Network Digital Twin}
\label{Convergence toward Optical Network Digital Twin}
Coherent transmission, QoT modeling based on physics models, and open optical networking have each evolved over different paths, but to deploy them in actual operation, a uniform software framework that can represent, evaluate, and optimize the entire optical network has been necessary.
This concept developed over a period of more than 20 years through engineering design and development of operation tools---it did not suddenly arise along with the recent appearance of the words ``Digital Twin.''
One of the early important milestones occurred in 2006 with a report on a method for explicitly incorporating transmission degradation in Routing and Wavelength Assignment (RWA) in a wavelength-routing optical network that takes impairment into account~\cite{Cardillo2006TransmissionImpairments}.
This abstraction established the important concept of expressing diverse physical-layer phenomena as a common QoT metric applicable to software optimization, which was later connected to GSNR-based analytical QoT models.
Then, in 2010, a design philosophy taking the physical layer into account was extended to the Elastic Optical Network through distance-adaptive spectrum resources~\cite{Jinno2010DistanceAdaptive}.
5
The next big step occurred in 2011 with the deployment of Network Planning and Operation Tool (NPOT) taking impairments into account~\cite{Azodolmolky2011NPOT}.
While previous design tools focused mainly on offline network design, NPOT integrated network design, QoT estimation, and operation support in a common software framework.
As optical networks became increasingly programmable and continuous telemetry collection from coherent TRxs and line systems became possible, the role of such software platforms expanded beyond offline design.
Instead of just evaluating network configuration beforehand, it became possible to continuously determine and model the state of the operational network and support network optimization based on that information.
One milestone typical of this transition is the idea of Near-Zero Margin Operation proposed by Boertjes \textit{et al.}
This research presented a framework for minimizing transmission margin and maximizing network capacity while grasping the state of the network during operation by combining SNR evaluation based on physical models, telemetry from coherent TRxs and line systems, centralized analysis/control, and closed-loop optimization~\cite{boertjes2019nearzero}.
In parallel with the evolution of operational software platforms, GNPy enabled open and vendor-neutral QoT estimation based on analytical physical-layer models.
At the same time, optical networking itself underwent a fundamental transformation through the stepwise disaggregation of both hardware and software.
In contrast to the embedding of monitoring, design, and control functions in vertically integrated network equipment, operators could now deploy new software components independent of base optical hardware using standardized interfaces.
The combined progress of operation workflows, open physical models, and software-defined network architectures enabled the digital representation of continuously synchronized optical networks, which laid the foundation for the evolution of past engineering design tools to today's ONDT~\cite{wang2022optical,akinrintoyo2023reconfigurable,raj2023digital,wang2024multispan}.
At the same time, advances in measurement-informed modeling~\cite{nishizawa2024field,ip2024inline,mano2023first}, transfer learning~\cite{raj2025generalized,raj2023self,wang2023transfer,xie2025wdm}, and AI-assisted physical-layer modeling improved the adaptability of digital models to heterogeneous devices, vendors, and operational environments, thereby enhancing the deployability of ONDT~\cite{wang2025multi,raj2024interference}.
More recently, these developments have evolved toward AI-agent-assisted autonomous optical networks, where ONDT serves as a trustworthy digital foundation for intelligent network operation~\cite{wang2026agentic}.
ONDT thus emerged not from a single technical breakthrough, but from the convergence of physical-layer modeling, monitoring, telemetry, open and programmable networking, and intelligent operation platforms.
\subsection{Architectural Design Dimensions for ONDT}
\label{Architectural Design Dimensions for ONDT}
As described in the previous section, ONDT emerged through the convergence of multiple technical layers. However, in current ONDT research, differences can be seen in the definition of Digital Twin as well as in which architectural aspects the research emphasizes. For example, some studies focus on the functional interaction between the physical network and virtual network, while others emphasize controller hierarchy, multi-domain collaboration, device-native monitoring, software implementation framework, etc.
To systematically organize these complementary perspectives, this survey characterizes ONDT architecture into the four design dimensions of Ecosystem Architecture, Functional Architecture, Control Architecture, and Implementation Architecture.
Here, Ecosystem Architecture identifies the stakeholders and organizational domains that create, exchange, and use digital representations. Functional Architecture defines the logical functions required to observe, represent, analyze, and operate the physical network.
The control architecture describes how decision and control responsibilities are assigned to operators, controllers, digital twins, and intelligent agents. Finally, the implementation architecture specifies how these capabilities are implemented using telemetry systems, databases, physical layer models, software platforms, and computing infrastructure.
These dimensions are complementary to each other and do not constitute a mutually exclusive hierarchy. The actual ONDT implementation is achieved by selecting and combining architectural options from all four dimensions according to the target operational scenario.
\subsubsection{Ecosystem Architecture}
\label{Ecosystem Architecture}
\begin{figure}[hb!]
    \label{ecoArchitecture}
  \centering
  \includegraphics[width=1.0\hsize]{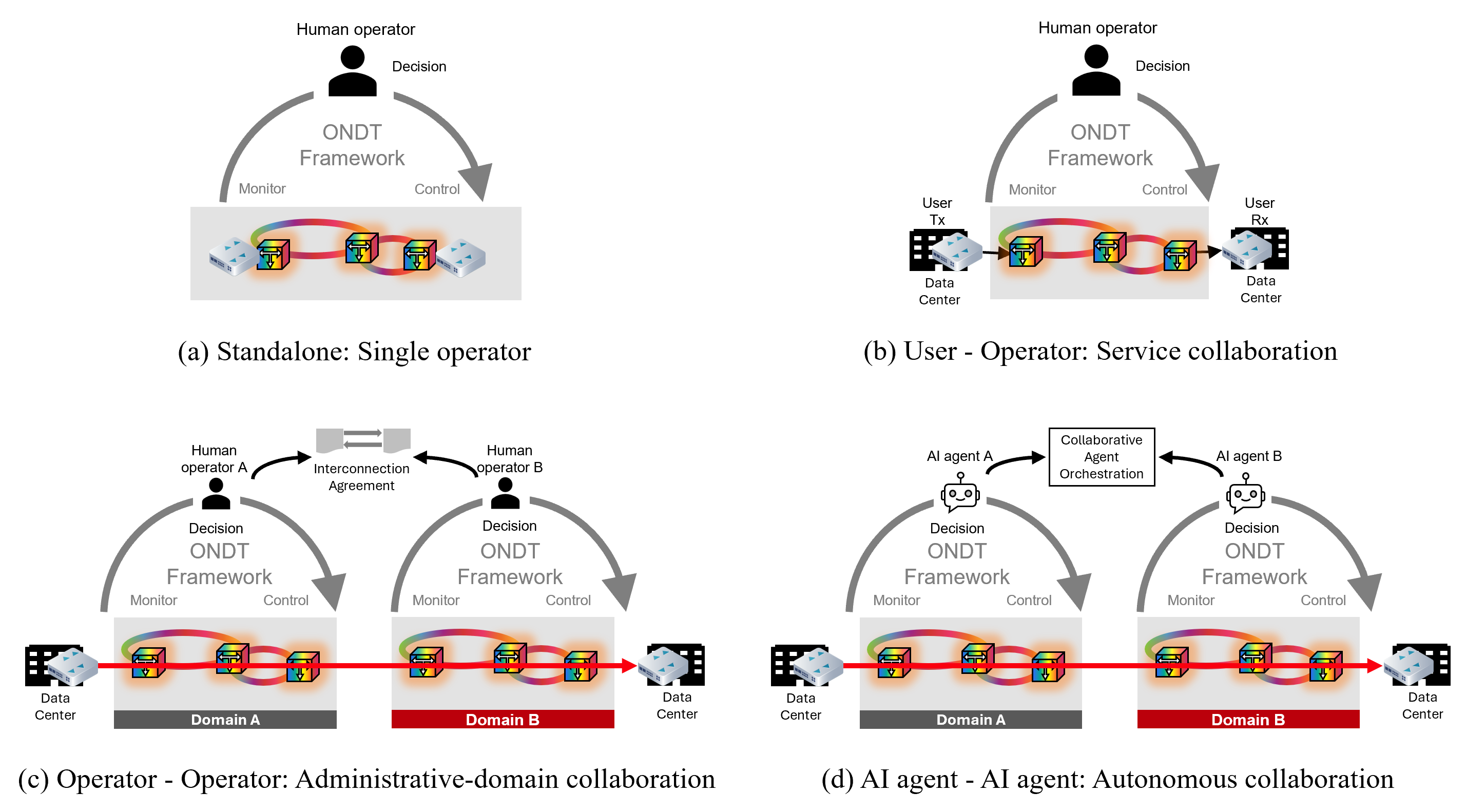}
 \caption{Typical examples of Ecosystem Architecture in ONDT.}
\label{fig:ecosystem_archi}
\end{figure}

Ecosystem architecture describes how stakeholders and administrative domains collaborate through ONDT, especially when E2E light paths cross multiple organizational boundaries.
Fig.~\ref{fig:ecosystem_archi} shows typical examples of Ecosystem Architecture from standalone operation to AI agent-assisted multi-domain collaboration.
In the standalone configuration (Fig.~\ref{fig:ecosystem_archi}(a)), a single operator maintains the ONDT and controls its physical network within a common administrative boundary.
In User–Operator Collaboration (Fig.~\ref{fig:ecosystem_archi}(b)), users operate terminals (i.e., coherent pluggable modules or transponders) while network operators provide line-system or spectrum connectivity ~\cite{Kaeval2021QoT,Nishizawa2023dynamic}. ONDT enables the exchange of information required for E2E QoT estimation without disclosing proprietary implementation details.
In multi-stakeholder ONDT ecosystems, however, infrastructure ownership, information ownership, information-access rights, and control authority do not necessarily coincide. Such scenarios require the Ecosystem Architecture to define not only what information is exchanged across organizational boundaries, but also who is authorized to access it, who can execute control actions, and how operational responsibility is distributed among the participating stakeholders ~\cite{Mitrovska2025Governance,Mitrovska2026Ownership}.
The configuration shown in Fig.~\ref{fig:ecosystem_archi}(c) is Operator–Operator Collaboration. Here, independent network operators collaborate without interfering with each other’s domain autonomy, thereby achieving E2E optical services that span multiple administrative domains ~\cite{Kaeval2025OSaaS,Nishizawa2025_ONDT}. Each operator discloses only abstracted information necessary for inter-domain operation while maintaining its own ONDT. An operational agreement defines information semantics, boundaries of responsibility, and validation procedures between operators ~\cite{Mitrovska25crossborder,zaid2026ecoc}.
Finally, Fig.~\ref{fig:ecosystem_archi}(d) shows a new Agentic Ecosystem in which AI agents act as proxies for users, network operators, and service providers. In this configuration, AI agents do not directly access the physical network. They perform inferences on their respective ONDTs and collaboratively negotiate candidates for operation actions spanning administrative domains.
Such interactions can support closed-loop Sense–Think–Act workflows, multi-agent coordination, and contract-driven orchestration while maintaining domain autonomy through explicit operational rules and the exchange of abstracted information ~\cite{Tanimura2026OFC,Ciena2025AgenticAI}. However, the execution of final network operation will continue to be managed by established operation policies, authorization mechanism, and controlled actuation interfaces.
\subsubsection{Functional Architecture}
\label{Functional Architecture}
\begin{figure}[hb!]
  \centering
  \includegraphics[width=0.8\hsize]{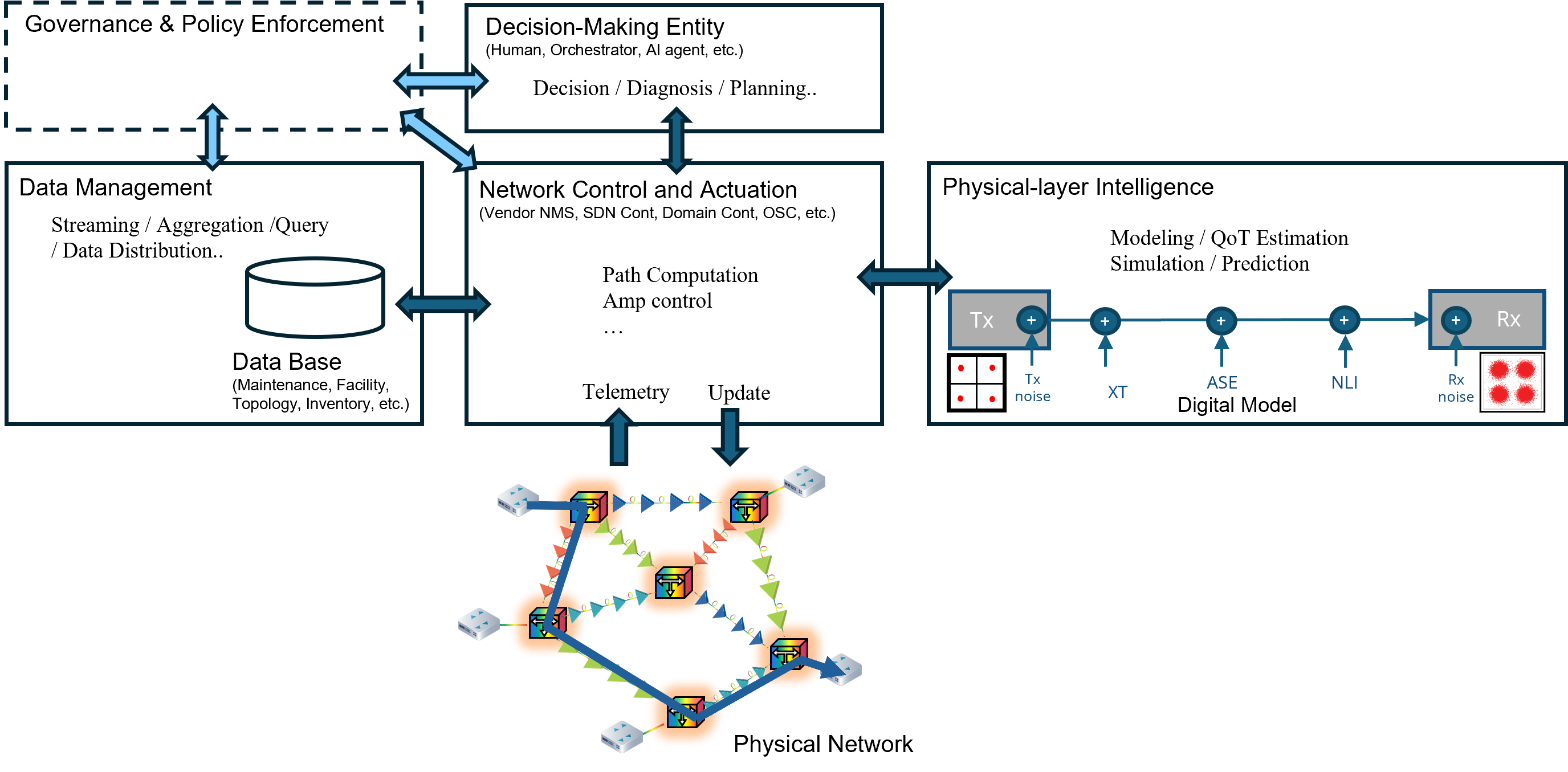}
\caption{Typical ONDT Functional Architecture.}
\label{fig:functional_archi}
\end{figure}

Functional Architecture defines the logical functions needed to maintain and use ONDT, independent of where these functions are implemented or which organization operates them. As shown in Fig.~\ref{fig:functional_archi}, ONDT forms a real–virtual feedback loop connecting the physical network, digital representations, physical-layer intelligence, decision-making, and network control and actuation.
Depending on network architecture and available measurement functions, information representing the current state of the physical network can be obtained from telemetry, alarms, coherent DSP, optical monitoring equipment, and operation databases.
Collected information is integrated into digital representations of the network. The digital model combines dynamic operation data with relatively static information such as topology and inventory stored in engineering and operation databases that must be accurate and consistent with the actual network to support reliable ONDT-based intelligent network operations. In addition to directly observable states such as receive power and pre-FEC BER, estimated physical-layer quantities can also be represented.
Depending on the deployment scenario, ONDTs may also incorporate Data Management and Governance and Policy Enforcement functions, as illustrated in Fig.~\ref{fig:functional_archi}. Data Management is responsible for the lifecycle of information assets used by the ONDT and serves as an abstraction layer that harmonizes diverse schemas, telemetry formats, and data models ~\cite{akbari2026ecoc}.
A complementary function is Governance and Policy Enforcement, which establishes the rules under which information and operational capabilities can be accessed, exchanged, and utilized ~\cite{Mitrovska2025Governance,Shariati23sovereingty}. Rather than functioning as a separate control entity, governance provides the trust, compliance, and accountability framework for interactions with ONDT resources and the physical network ~\cite{Mitrovska2026Ownership,Mitrovska25crossborder}.
Digital representations are processed by physical-layer intelligence functions, which integrate analysis models, measurement-informed models, and learning-based approaches to estimate and predict network behavior. One or more complementary modeling techniques can be used depending on the target application and available information ~\cite{Zhuge2023BuildingDTOptical,Faruk2024MeasurementInformedDT,Zhuge2025DigitalTwinAIAgent,Nishizawa2026JOCN}.
Typical functions include QoT estimation, anomaly detection, fault localization, parameter calibration, and evaluation of candidate network configurations.
Obtained information is provided to a decision-making entity, which may be a human operator, deterministic optimization engine, service orchestrator, or AI agent. Based on the estimated network state and operation targets, an appropriate operation action is selected while taking operational constraints into account.
Approved decision-making is converted to network settings by the network control and actuation function. Resulting changes in the network are then observed by telemetry and monitoring, and the digital representations are updated, closing the real–virtual feedback loop.
In conventional optical network architecture, it is common to provide design tools and management systems as separate products. This Functional Architecture, however, explicitly defines a physical-layer intelligence layer that provides a bridge between digital representations and decision-making operations.
Such functional decomposition enables the independent evolution and implementation of design tools, control equipment, and management systems and facilitates interoperability and deployability.
The Functional Architecture shown here functions as a framework for organizing the enabling technologies discussed in the next section.
\subsubsection{Control Architecture}
\label{Control Architecture}
\begin{figure}[hb!]
  \centering
  \includegraphics[width=0.8\hsize]{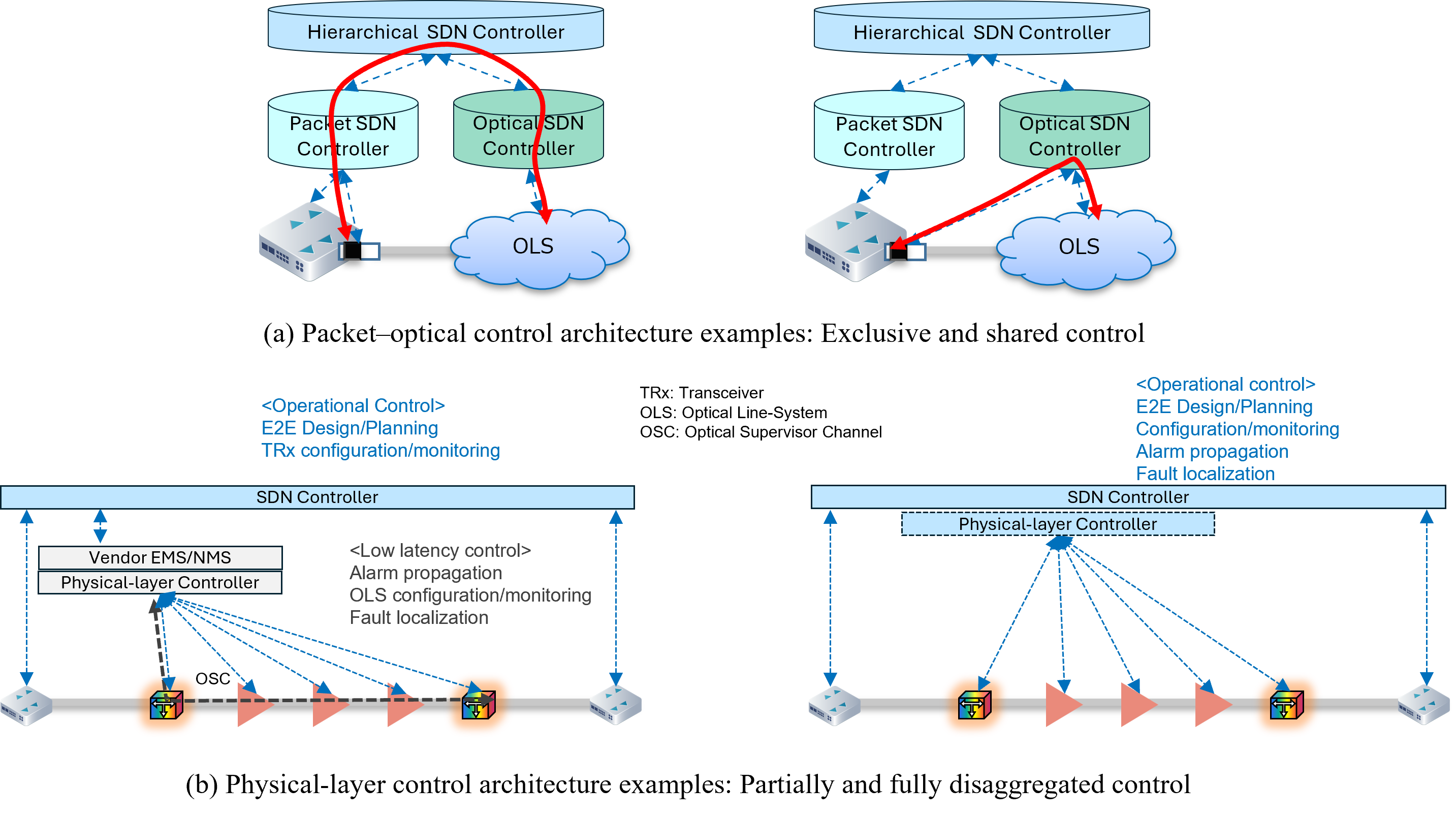}
\caption{Typical ONDT control architecture examples.}
\label{fig:control_archi}
\end{figure}

Control Architecture describes how operational responsibility is divided among network elements, controllers, ONDT platforms, human operators, and AI agents. It therefore concerns where control authority is placed and how responsibilities are coordinated across different network layers and control domains. As illustrated in Fig.~\ref{fig:control_archi}, two complementary aspects need to be considered in optical networks: packet–optical control and physical-layer control.
Coherent pluggable TRxs introduce a unique control challenge in IPoWDM networks because the TRx is physically hosted inside the IP router while its optical operating parameters directly affect transmission performance across the OLS. Consequently, the information required to configure the device is distributed between packet and optical domains. The IP SDN controller is responsible for service activation and packet-layer configuration, while the optical SDN controller possesses the physical-layer intelligence required to determine feasible optical operating parameters.
Fig.~\ref{fig:control_archi}(a) shows representative packet–optical control architectures. Two approaches have been demonstrated for controlling packet–optical nodes equipped with coherent pluggable TRxs: exclusive and shared control ~\cite{Giorgetti2023Hierarchical}. In the exclusive approach, the packet SDN controller provides exclusive access to the packet optical nodes. On the other hand, the optical parameters determined by the optical controller are propagated through the hierarchical and packet controllers. In the shared approach, the packet and optical SDN controllers directly manage their respective functions within the same packet optical node.
More advanced architectures enable governed access, where both controllers can interact with the pluggable under predefined authorization policies. In such approaches, ownership, operational responsibility, and control authority are explicitly separated, allowing the optical SDN controller to configure optical attributes while the IP SDN controller retains authority over packet-layer functions and service activation ~\cite{Mitrovska2026Ownership}.
An emerging ONDT concept has been discussed as a physical-impairment-aware control-plane element that interacts with the optical controller and provides feedback on transmission parameters ~\cite{Giorgetti2023Hierarchical}.
Fig.~\ref{fig:control_archi}(b) shows two representative configurations of physical-layer control: partially disaggregated and fully disaggregated. In the partially disaggregated configuration, higher-level operational functions can be executed through an open SDN controller or service orchestrator. In contrast, line-system functions requiring low-latency and deterministic operation can remain within the vendor EMS/NMS and physical-layer controller. Microsoft’s cloud optical network architecture provides a representative example ~\cite{Filer2016Microsoft}.
Leaving selected control functions within a vendor system is not necessarily a constraint on ONDT. Consequently, practical brownfield deployments can adopt a hierarchical and partially disaggregated control architecture, in which standardized interfaces are introduced for higher-layer orchestration and service provisioning, while selected lower-layer functions remain under vendor-specific control.
In the fully disaggregated configuration, a common physical-layer controller assumes a broader range of configuration, monitoring, alarm, and fault-management functions across optical devices. This architecture can improve implementation flexibility and vendor interoperability, but requires more comprehensive information models, standardized interfaces, operational validation, and clear boundaries of responsibility.
These two architectural configurations should not be understood as a step toward full granularity, but as alternative design options for how operational responsibilities are assigned. In short, real-world ONDT deployments can adopt different types of control architectures to achieve the right balance between interoperability, scalability, operational efficiency, and deployability.
Here, the ONDT itself is not a control entity; it supports operational decisions and provides state estimation, prediction, and evaluation of candidate actions. Depending on the adopted control architecture, execution authority is delegated to a specific control entity, such as a human operator, an SDN controller, or an AI-assisted orchestration system. This separation allows the same ONDT to be used for manual operations, deterministic automation, or AI agent-assisted operations without fundamentally changing the underlying network control architecture. 
This separation of physical layer control and service-level control is also effective when deploying ONDT functionality incrementally.
\subsubsection{Implementation Architecture}
\label{Implementation Architecture}
Implementation Architecture describes how the Ecosystem, Functional, and Control architectures are embodied using physical devices, software platforms, and computing infrastructures. In contrast to Functional Architecture, Implementation Architecture deals with where each ONDT function is to be executed, how information is to be stored and exchanged, and how software components are to be integrated.
\begin{figure}[hb!]
  \centering
  \includegraphics[width=0.8\hsize]{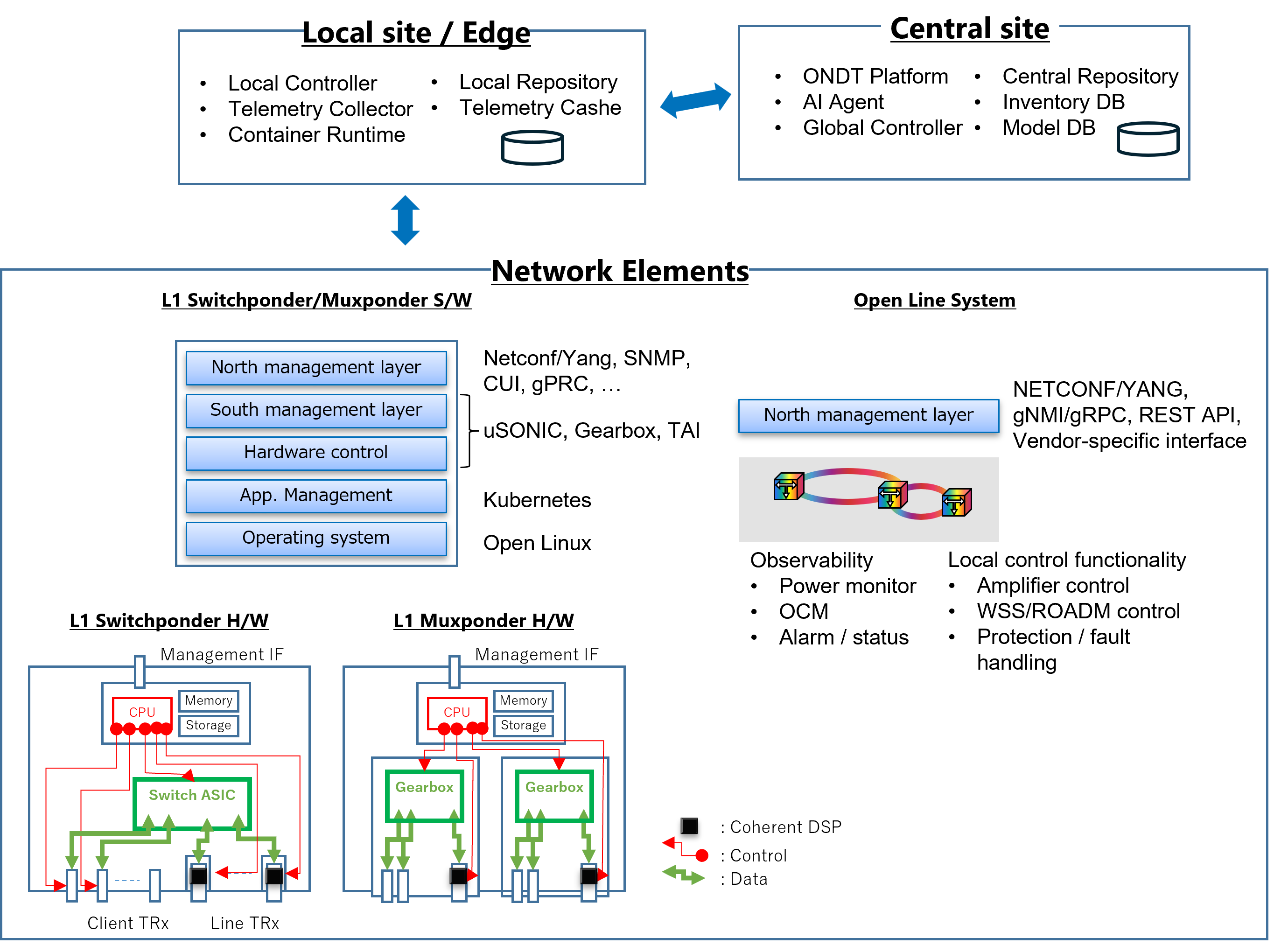}
\caption{Representative Function Placement for ONDT Implementation}
\label{fig:implementation_archi}
\end{figure}
Fig.~\ref{fig:implementation_archi} shows a representative function placement for ONDT implementation. At the device level, network elements like transponders and line-systems not only obtain telemetry information but also perform intra-device processing before disclosing the information to upper-level platforms.
By executing these functions locally, network devices can provide standardized or abstracted information to upper-level ONDT applications while reducing their communications overhead and maintaining proprietary implementation details.
An edge platform consolidates telemetry from nearby network elements, executes operation functions that take local data processing and physical models into account, and selectively exchanges only abstracted information with a centralized platform. This scheme can reduce communications overhead and support practical operation workflows ~\cite{Shen2025UnifiedTelemetry,Nishizawa2024semi-automatic}.
A container runtime provides a portable software execution environment for ONDT applications placed at the edge or on a centralized platform. This enables software components to be independently placed, updated, scaled, and migrated without having to change the base hardware platform.
Such execution environments make it possible to place the same ONDT software components in devices, at the edge, or on a centralized platform.
The centralized platform generally executes computationally intensive ONDT functions such as digital repository, physical-layer model, global optimization engine, design applications, and AI agent. These functions integrate multiple network elements and information collected from edge platforms and support overall network analysis, prediction, and decision-making.
Advances in open optical networking and cloud-native software architecture have enabled these software components to be flexibly placed in coherent transponders, edge servers, domain controllers, and cloud platforms. This flexibility enables ONDT functions to be deployed in a stepwise manner while maintaining existing operation systems and vendor-specific control mechanisms.
The typical functional arrangement shown in Fig.~\ref{fig:implementation_archi} does not specify the owners of individual software components. In Ecosystem Architecture, equivalent ONDT functions can be deployed by different stakeholders.
There is no universally optimal Implementation Architecture. Practical ONDT deployments range from tightly coupled hybrid implementations to cloud-native modular architectures composed of independently deployable services ~\cite{Nishizawa2026JOCN,Nishizawa2024semi-automatic}. The most appropriate implementation depends on deployment scale, latency requirements, operational constraints, confidentiality, and the degree of openness required.
The practical characteristics of ONDT are determined by these four architectural dimensions. Ecosystem Architecture affects information availability, confidentiality, and interoperability between closed domains. Functional Architecture determines which physical states to represent and how to form a real-virtual feedback loop. Control Architecture determines operational authority, response times, and security boundaries. Finally, Implementation Architecture determines the computational complexity, interface requirements, and migration man-hours of the resulting system.
These four architectural dimensions must be evaluated in a comprehensive manner rather than on an individual basis. Although detailed Functional Architecture has the potential of improving physical accuracy, it requires more information sharing and implementation complexity. Distributed control can reduce response latency, but it can complicate the maintenance of consistency and governance. An open and modularized implementation can simplify a migration, but vendor-specific functions can provide higher accuracy and reliability at the device level.
Taking the above into account, we introduce FDD in the next section as a common framework for evaluating the trade-offs among these architectural dimensions.
\subsection{Go-To-Market Requirements}
\label{Go-To-Market Requirements}
In this paper, we introduce the three basic dimensions of FDD to systematically characterize an ONDT system and its practical operational requirements.
%
\subsubsection{Fidelity}
\label{subsubsec:Fidelity}
Fidelity refers to the ability of an ONDT system to accurately reproduce and estimate the state of the physical layer and transmission behavior. Fidelity is therefore not simply a desirable characteristic but rather a basic prerequisite to reliable Digital Twin operation. In optical networks, however, transmission degradation originates in complex physical phenomena such as nonlinear propagation, wavelength-dependent device characteristics, and dynamic operation conditions, so satisfying this requirement is particularly difficult. We can therefore say that Fidelity in ONDT is the ability to continuously maintain physically consistent representations to support highly reliable predictions, optimization, and autonomous operation.
\subsubsection{Dynamicity}
\label{subsubsec:Dynamicity}
The second basic requirement of ONDT is Dynamicity, which is the ability of the Digital Twin to maintain synchronization with an ever-changing physical network.
While Fidelity addresses how accurately a physical network can be represented, Dynamicity addresses how well a digital representation can be updated according to operational requirements. Therefore, Dynamicity should not be interpreted as simply maximizing the frequency of telemetry updates, but as maintaining cyber-physical consistency at a time scale that is appropriate for the target application.
The required synchronization frequency varies greatly from application to application. While network design and capacity planning may require occasional updates of network parameters, service provisioning requires synchronization at the time scale of connection establishment. This requirement is particularly important in optical transport networks, where the physical state changes at multiple time scales. As shown in Fig.~\ref{fig:dt_application_mapping}, ONDT must balance synchronization cost, computational complexity, and deployability to achieve the level of dynamicity required for the desired behavior.
\subsubsection{Deployability}
\label{subsubsec:Deployability}
In many actual production environments, optical transport networks consist of multiple vendors and generations of devices, with long infrastructure lifecycles and established operational workflows. Here, practical ONDTs must be designed with Deployability in mind, in addition to Fidelity and Dynamicity. Deployability means the ability to deploy and operate ONDTs under technical, operational, and economic constraints while maintaining interoperability, scalability, and long-term serviceability. It goes beyond mere "implementability" as it requires coexistence with existing network management systems, operational workflows, and partially disaggregated systems. It enables phased deployment while minimizing operational impact and maximizing infrastructure reuse.
Deployability must be considered with Fidelity and Dynamicity. Often, higher Fidelity requires more sophisticated physical models, denser telemetry, and computational complexity, while higher Dynamicity requires more frequent synchronization and communication overhead. Therefore, the objective of ONDT is not to maximize Fidelity or Dynamicity separately, but to balance all three aspects according to the target operational scenario and ensure sufficient operational value within acceptable engineering effort, operational impact, and investment costs.
In addition to the technical requirements of FDD, practical ONDT systems require transparency, traceability, and visualization so that network operators can recognize, verify, and confirm estimated network states, QoT estimates, and automated control actions. ONDT systems must maintain robust operations across different network environments, different vendor implementations, and changing operational conditions.
Additionally, governance mechanisms are required to define how telemetry, models, derived knowledge, and operational recommendations are collected, shared, updated, and validated across organizational and vendor boundaries.
\section{Enabling Technologies}
\label{Enabling Technologies}
Based on the conceptual framework and FDD requirements
established in the previous section, we examine the
enabling technologies of ONDT and their tradeoffs
for practical deployment.
\subsection{Modeling}
\label{Modeling}
\subsubsection{TRx Modeling}
\label{TRx Modeling}

\begin{table*}[t]
\centering
\caption{TRx-model selection for ONDT applications under the FDD framework.}
\label{tab:trx_models_fdd}
\label{tab:trx_equalizer_models}
\scriptsize
\setlength{\tabcolsep}{1.5pt}
\renewcommand{\arraystretch}{1.10}
\begingroup
\newcommand{\trxcell}[1]{\raggedright #1\par}
\newcommand{\trxlevel}[1]{\centering #1\par}
\begin{tabular}{|p{0.19\linewidth}|p{0.075\linewidth}|p{0.09\linewidth}|p{0.105\linewidth}|p{0.27\linewidth}|p{0.22\linewidth}|}
\hline
\trxcell{\textbf{TRx model and core abstraction}} &
\trxlevel{\textbf{Fidelity}} &
\trxlevel{\textbf{Dynamicity}} &
\trxlevel{\textbf{Deployability}} &
\trxcell{\textbf{Required data or measurements}} &
\trxcell{\textbf{Primary ONDT role}}
\tabularnewline
\hline

\trxcell{\textbf{Mode-specific threshold.}
Margin against required OSNR/GSNR.} &
\trxlevel{Low} &
\trxlevel{Low} &
\trxlevel{Very high} &
\trxcell{Mode catalog; required OSNR/GSNR; static penalty masks.} &
\trxcell{Network-wide feasibility screening and admission control.}
\tabularnewline
\hline

\trxcell{\textbf{B2B equivalent SNR.}
Per-unit $\mathrm{SNR}_{\mathrm{TRx}}(\mathrm{ROP})$ and inverse-SNR aggregation.} &
\trxlevel{Medium} &
\trxlevel{Medium--high} &
\trxlevel{High} &
\trxcell{Per-unit ROP sweep and BER/Q mapping; preferably OSNR sweep; fitted $N,D$ and calibration range.} &
\trxcell{Routine QoT estimation, module-spread representation, and assurance.}
\tabularnewline
\hline

\trxcell{\textbf{Coupled ZFE.}
Frequency-domain channel inversion.} &
\trxlevel{Medium--high} &
\trxlevel{High} &
\trxlevel{Medium--high} &
\trxcell{B2B calibration; complex WSS/AA responses; signal spectrum; ASE powers and injection locations.} &
\trxcell{Fast filtering-aware estimation and impairment disaggregation.}
\tabularnewline
\hline

\trxcell{\textbf{Coupled linear equalizers.}
MMSE/FSE address the ISI/noise tradeoff; FLE/WFLE capture finite DSP memory.} &
\trxlevel{High--very high} &
\trxlevel{Medium--high} &
\trxlevel{Low--medium} &
\trxcell{ZFE inputs; signal/noise spectra or covariances; sampling rate. Finite-tap models also require tap count, receiver bandwidth, and per-unit validation; whitening response for WFLE.} &
\trxcell{Filtering-sensitive and low-margin paths; finite-memory assessment, calibration, and benchmarking.}
\tabularnewline
\hline
\end{tabular}
\endgroup
\end{table*}

An ONDT cannot infer E2E QoT from fiber and amplifier models alone. The TRx closes the modeling chain by translating the physical state of the optical path into the QoT metric used for planning, provisioning, or assurance. TRx impairments also affect the impact of NLI and WSS modeling inaccuracies on E2E QoT, as discussed in Sections~\ref{From Digital Coherent Reception to GN Modeling} and~\ref{WSS Modeling}, respectively. Here, we examine the main TRx approaches within the FDD framework and discuss their implications for ONDT design purposes. 

The lowest-complexity abstraction is a mode-specific receiver threshold, used to determine whether a given lightpath is feasible. This approach is adopted in operational-mode specifications such as OpenROADM and OpenZR+ \cite{OpenROADMOpticalSpecV9,OpenZRPlusSpecV3}. %
Although rapid and readily deployable, a threshold cannot predict QoT or represent received optical power (ROP) dependence and WSS and DSP interaction. 
Measurement-based implementation-penalty models and open QoT estimators have therefore progressively introduced explicit transponder terms into physical-layer models \cite{Curri2022GNPyPhysicalLayer,Wang2024DigitalTwinOpticalReview,Kaliteevskiy2020TransponderPenalty,Filer2018OpenQoT}.
A practical next level is a per-unit back-to-back (B2B) equivalent-SNR model. External optical noise is removed or controlled, the ROP is swept with a variable optical attenuator, and the measured BER or Q-factor is mapped to an equivalent TRx SNR. A compact saturating law for the ROP dependence is provided in \cite{Mano2024TRXROP}, and further exploited for filtering penalty assessment in \cite{miotto_generalized_2026}. Under a scalar Gaussian-channel abstraction, this contribution can be combined with line noise through
\begin{equation}
\mathrm{SNR}^{-1} \simeq
\mathrm{SNR}_{\mathrm{ASE}}^{-1}
+\mathrm{SNR}_{\mathrm{NLI}}^{-1}
+\mathrm{SNR}_{\mathrm{TRx}}^{-1},
\label{eq:trx_inverse_snr}
\end{equation}
where all terms must use a consistent signal and noise-bandwidth convention, and the NLI term is omitted in an effectively linear regime. The combined SNR can then be mapped to BER or Q-factor using the calibrated transmission-mode relationship.

The scalar inverse-SNR approximation in \eqref{eq:trx_inverse_snr} is not universally sufficient, particularly under strong cascaded-WSS filtering, where distributed ASE, receiver noise, and DSP equalization interact. In such cases, filtering and TRx impairments must be evaluated jointly rather than as independent static penalties (see Sec.~\ref{WSS Modeling}) \cite{miotto_generalized_2026,torres-ferrera_filtering_2024}.
A filtering-aware TRx model therefore represents the coherent receiver as an optical front end followed by an equivalent receiver-noise source, an analog anti-aliasing (AA) response, sampling and analog-to-digital conversion, and digital equalization. 
With only E2E measurements, transmitter and receiver penalties can be fitted jointly. When justified, a receiver-dominated approximation places equivalent TRx noise at the receiver input, preserving explicit line filtering while capturing proprietary effects of commercial devices through a measurement-derived receiver model \cite{Kaliteevskiy2020TransponderPenalty,miotto_generalized_2026}.
The equivalent TRx noise may not be spectrally white. The model proposed in \cite{miotto_generalized_2026} also accounts for electrical-bandwidth limitations and front-end response, while allowing the effective noise power spectral density to be inferred from measurements rather than imposed from vendor-internal specifications.
At the highest fidelity, the filtered signal and noise contributions are jointly evaluated after receiver equalization, accounting for their distinct spectral characteristics. The WSS-dependent propagation of distributed ASE is discussed in Section~\ref{WSS Modeling}.

The equalizer abstraction sets the tradeoff between residual inter-symbol-interference (ISI) and noise enhancement. A zero-forcing equalizer (ZFE) inverts the equivalent channel wherever inversion is well conditioned, providing a simple analytical reference and enabling disaggregation with respect to individual noise sources and line elements, but it may strongly amplify noise near spectral notches. A minimum-mean-square-error equalizer (MMSEE) instead uses the signal and noise statistics to balance residual ISI against noise enhancement. A fractionally spaced equalizer (FSE) represents the sampled receiver with more than one sample per symbol, avoiding an idealized analog matched-filter assumption and capturing sampling-domain effects more faithfully \cite{miotto_generalized_2026,Savory2008DigitalFilters}. These labels describe complementary modeling choices: ZF and MMSE specify an optimization criterion, whereas symbol-spaced or fractionally spaced operation specifies the sampled representation.
Real coherent receivers additionally impose a finite-memory constraint. A whitened finite-length equalizer (WFLE) retains a convenient matrix formulation after idealized noise whitening. A non-whitened finite-length equalizer (FLE) instead constructs the sampled colored-noise covariance directly. The latter is more faithful when the AA response, colored TRx noise, or tight and asymmetric optical filtering is relevant, because it avoids the long temporal tails and truncation sensitivity introduced by ideal whitening. This gain in fidelity requires a denser covariance construction and matrix solution, as well as calibration of the effective tap count and receiver bandwidth \cite{miotto_generalized_2026,parisi_analytical_2025,torres-ferrera_impact_2025}.

Model selection should be tied to the ONDT function through FDD. Here, fidelity denotes the physical effects resolved within the calibration domain; dynamicity denotes the ability to update the estimate with operating state and telemetry; and deployability reflects the availability of measurements, model parameters, and computational resources. Table~\ref{tab:trx_models_fdd} organizes the complete TRx hierarchy on these axes. 

\subsubsection{Fiber Propagation Modeling}
\label{Fiber Propagation Modeling}
%
%
%
%
%

Among the fiber-propagation effects, the key contributions for channel-resolved QoT estimation in modern coherent WDM systems are undoubtedly the Kerr effect, generating NLI, and SRS.
%
%
%
%

For SRS modeling, the main requirement is the accurate evaluation of the WDM power evolution along both frequency and propagation distance~\cite{raj2023machine}. 
This is particularly important in multiband systems, where the frequency dependence of fiber loss, CD, and effective area must be included, and where the inter-band SRS-induced power transfer can become the leading interaction among bands~\cite{damicoInterband,Correia:21}. 
SRS is described by a set of coupled ordinary differential equations, which can be solved numerically~\cite{AGRAWAL2013,bromage2004raman,tariq1993} or approximated under simplifying assumptions~\cite{chraplyvy1984optical, LasagniRaman}. 
More recently, perturbative solutions have been introduced to provide a controllable accuracy: the target precision can be obtained by selecting the appropriate expansion order, with second- and third-order truncations shown to be sufficient in the considered C+L+S and U-to-E scenarios, respectively~\cite{d2023introducing}.

For Kerr-induced NLI modeling, the reference physical description is given by the nonlinear Schrödinger equation or, in dual-polarization coherent systems, by the Manakov equation~\cite{AGRAWAL2013,menyuk-interaction_pmd-jlt-2006}. 
Although split-step-fourier-method (SSFM) remains the main waveform-level reference, its computational cost is not suitable for repeated path computation, power optimization, or ONDT synchronization. 
Analytical and semi-analytical models therefore provide the practical modeling layer. 
Early GN-model formulations introduced the approximation of treating the accumulated Kerr distortion as an additive Gaussian noise contribution, providing a practical accuracy vs complexity trade-off for system-performance prediction~\cite{Poggiolini2012GNModel,Carena2012ModelingNLI,poggiolini2014jlt}.

Related derivations and network-oriented formulations further clarified the model assumptions, provided closed-form or faster numerical implementations, and adapted GN-based estimation to flexible-grid scenarios with non-identical channel frequencies, bandwidths, and launch powers~\cite{Poggiolini_GN_Derivation,Johannisson_Agrell_Network_GN}.
%
%
%
As introduced in Sec.~\ref{From Digital Coherent Reception to GN Modeling}, the GN model~\cite{Poggiolini_GN_Derivation}, EGN model~\cite{Carena:14}, and the approach developed by Dar~\cite{Dar:13, Dar:14}, provide different descriptions of nonlinear interference, while Raman-aware and generalized formulations, including ISRS-GN, GGN, and related scalable or disaggregated approximations, extend this modeling family to wideband conditions by accounting for non-uniform power evolution and frequency-dependent fiber parameters~\cite{Cantono2018GGN,semrau2020modeling,LasagniRaman,damicoJLT2022}.

\begin{table*}[!ht]
\centering
\caption{Recommended GN-family modeling levels for representative ONDT functions and transmission regimes.}
\label{tab:fiber_model_selection}
\begingroup
\addtolength{\tabcolsep}{-0.6\arrayrulewidth}
\resizebox{\textwidth}{!}{%
\begin{tabular}{
|>{\raggedright\rightskip=0pt plus 2em\arraybackslash}p{2.8cm}
|>{\raggedright\rightskip=0pt plus 2em\arraybackslash}p{4.4cm}
|>{\raggedright\rightskip=0pt plus 2em\arraybackslash}p{4.4cm}
|>{\raggedright\rightskip=0pt plus 2em\arraybackslash}p{4.4cm}
|>{\raggedright\rightskip=0pt plus 2em\arraybackslash}p{4.5cm}|}
\hline
\textbf{Transmission regime} &
\textbf{System design} &
\textbf{Lightpath feasibility} &
\textbf{Operational deployment} &
\textbf{Main FDD consideration} \\
\hline

Ecosystem Architecture: Fig.~\ref{fig:ecosystem_archi}(a) / fixed-route system &
EGN for high NLI accuracy; SSFM for detailed waveform-level validation &
EGN for few fixed paths with known co-propagating channels &
EGN with a slowly varying and comprehensive E2E system parameters, available in standalone architectures &
Route/channel aggregation acceptable for fixed E2E systems; additional complexity traded for higher Fidelity \\
\hline

Ecosystem Architecture: Fig.~\ref{fig:ecosystem_archi}(b) and Fig.~\ref{fig:ecosystem_archi}(c) / Disaggregated System &
GN for routine design; SSFM for detailed waveform-level validation &
GN with fast, span-by-span disaggregation &
GN-based QoT estimation; open implementations such as GNPy, based on common metrics and standard parameters &
Well-validated FDD balance; additional complexity only for decision-relevant NLI corrections \\
\hline

Short-reach &
GN usually sufficient for launch-power optimization as the TRx SNR degradation is dominant &
GN plus an appropriate TRx model; TRx impairments may dominate E2E QoT &
GN-based estimation with measured or calibrated TRx characteristics &
Prioritize E2E QoT accuracy; TRx characterization may matter more than NLI-model refinement \\
\hline

Multiband &
GGN for general wideband design; ISRS-GN within valid fiber-parameter and SRS assumptions &
Scalable or disaggregated GGN; ISRS-GN where sufficiently accurate &
GGN-based open QoT engines accounting for frequency-dependent fiber parameters and SRS &
Frequency-dependent power evolution for Fidelity; scalable, disaggregated computation for network applications \\

\hline
\end{tabular}%
}
\endgroup
\end{table*}

From an ONDT perspective, these modeling alternatives should not be interpreted as a sequence in which increasing physical detail is always preferable. 
The appropriate modeling level depends both on the transmission regime and on the target ONDT function. 
Fidelity must therefore be assessed with respect to the accuracy required by the final QoT estimation or optimization task. 
Dynamicity depends on whether both the model execution and the required physical state can be updated at the time scale of that task. 
Deployability additionally accounts for computational complexity, parameter availability, disaggregability, validation maturity, and compatibility with operational network environments.


The GN model is experimentally validated for dispersion-uncompensated long-haul links with spans substantially exceeding the effective length, where NLI predictions are generally accurate and slightly conservative~\cite{Torrengo:11,Nespola2014ptl}. Its efficient closed-form solutions and span-wise formulation support repeated route evaluations without a complete E2E description, making it a \emph{de facto} standard for lightpath feasibility and network-wide optimization in conventional C-band systems.
Neglected statistical correlations can, however, cause large, typically conservative errors in short-reach systems, although dominant TRx noise often reduces their practical impact~\cite{Mano2022ECOC}. 
The EGN model provides higher NLI prediction fidelity at greater computational cost and requires a detailed description of the complete route and co-propagating channels. These requirements limit the suitability of EGN and its extensions for multi-domain, dynamically routed meshed-network digital twins, which require fast, disaggregated evaluations. Their aggregated formulation is better suited to point-to-point digital twins, where the complete E2E configuration is available. Higher-fidelity models are also attractive for offline design, where relaxed timing constraints can justify additional complexity when accuracy materially affects the design outcome.

Short-reach experiments showed that GN-based estimation with appropriate TRx characterization can support launch-power optimization~\cite{Mano2022ECOC,DAmico:26GNPy}. In a three-span system, a 1-dB error in TRx SNR changed the Q-factor by 0.84~dB, compared with 0.08~dB for the same error in ASE or NLI SNR~\cite{Mano2023_TRxModel,Mano2022ECOC}. Thus, when TRx impairments dominate QoT, accurate TRx characterization can be more relevant than refining the NLI model.

%
%
%

The appropriate modeling hierarchy changes when the occupied optical bandwidth is extended beyond conventional C-band transmission. 
In this case, the Fidelity of the NLI estimate depends not only on the treatment of signal statistics, but also on an accurate description of the frequency-dependent fiber parameters and of the SRS-induced power evolution.
%
%
%
The GGN formulation addresses this limitation by retaining the frequency dependence of the relevant fiber parameters and the general SRS power evolution. 
ISRS-GN instead provides a simplified Raman-aware formulation under a more restricted set of assumptions.
The latter approximation allows the derivation of a closed-form expression that can generally be effectively used in the C band to include the SRS effect in the GN model, and in C+L scenarios when ultra-low-loss (ULL) fibers are used. When the occupied spectrum exceeds the C+L bands, the general GGN-model approach must be used, since, in addition to the flat-loss assumption, the triangular approximation of the SRS profile also introduces large inaccuracies \cite{Souza:23}.
From the FDD perspective, ISRS-GN therefore provides an attractive solution when its validity assumptions are satisfied and computational efficiency is a primary requirement. 
The GGN model provides a more general high-fidelity description for wideband systems, while scalable and disaggregated GGN approximations are particularly important when the same physical accuracy must be retained for repeated network-level calculations.
Furthermore, the GGN model is implemented in GNPy and has been extensively tested in greenfield and brownfield commercial scenarios \cite{Ferrari2020GNPy}, \cite{ferrari2020ecoc}, \cite{DAmico:22OrangeEXP}, \cite{damico2024icton}, \cite{DAmico:22}.

%
%


NLI modeling runtimes span several orders of magnitude: SSFM propagates waveforms numerically and can require minutes to hours for realistic WDM scenarios, while GN-family models avoid waveform propagation but may require frequency integrations and Raman-aware power-profile calculations, typically taking seconds to minutes.
Closed-form expressions and scalable approximations can reduce evaluation to tens of milliseconds. This enables interactive path calculations and highly dynamic ONDT applications. These times are indicative and vary depending on system parameters, numerical settings, and software/hardware implementation.

%
%
%
%
%
%

The difference between theoretical model accuracy and practical Deployability is particularly noticeable when GN-based models are integrated into an open QoS estimation framework and evaluated in an operational multi-vendor network.
%
Experimental validation of an open GN-based QoT estimator reported E2E SNR errors generally within $\pm0.75$~dB over paths up to 1,945~km with commercial multi-vendor equipment and evaluation times of milliseconds~\cite{Filer2018OpenQoT}. Subsequent GNPy validation reported OSNR and GSNR errors within 1~dB in more than 90\% of over 500 experimental conditions~\cite{Ferrari2020GNPy}. However, conservative NLI estimates do not guarantee conservative E2E QoT predictions: parameter and measurement uncertainties or incomplete Raman modeling can produce optimistic estimates~\cite{Filer2018OpenQoT,Ferrari2020GNPy}. Prediction errors should also be assessed against the network's temporal QoT variability~\cite{nishizawa2024fast}.
Coherent probing in production networks also supports GSNR-based QoT assessment across different TRx configurations~\cite{Kaeval2021QoT}. Strong filtering may nevertheless require separate penalty evaluation or receiver-specific modeling beyond simple noise accumulation (see Sec. \ref{WSS Modeling}).
The evolution of GNPy toward operational network control was further demonstrated through its integration with the OpenDaylight TransportPCE controller. Leveraging OpenROADM models and specifications, TransportPCE uses GNPy as an external QoT engine for vendor-neutral, impairment-aware path computation and optical-path validation~\cite{TransportPCE, Birk:20, LeRouzic:22}.
These results show that Deployability is not determined by computational cost alone. A practical ONDT model must rely on physical parameters that can be made available in operational environments, support heterogeneous network elements and multi-vendor transmission systems, and expose an interface compatible with network planning and control. In this respect, span-wise and disaggregated GN-based approaches provide a favorable compromise between physical fidelity and the requirements of operational network-level computation.

Table~\ref{tab:fiber_model_selection} summarizes an application-dependent strategy: offline design permits higher-fidelity models, while lightpath feasibility prioritizes efficiency and span-wise disaggregation.
Deployment also requires parameter availability, calibration, interoperability, validation, and network synchronization, so model selection should balance FDD rather than maximize physical accuracy alone.

%
%
%
%
%

\subsubsection{Optical Amplifier}
\label{Optical Amplifier Modeling}

\begin{table*}[!t]
\centering
\caption{Qualitative FDD assessment of optical amplifier modeling approaches.}
\label{tab:optical_amplifier_fdd}
\scriptsize
\setlength{\tabcolsep}{4pt}
\renewcommand{\arraystretch}{1.12}
\begin{tabular}{
|>{\raggedright\arraybackslash}p{0.21\textwidth}
|>{\raggedright\arraybackslash}p{0.26\textwidth}
|>{\raggedright\arraybackslash}p{0.25\textwidth}
|>{\raggedright\arraybackslash}p{0.23\textwidth}|}
\hline
\textbf{Modeling approach} &
\textbf{Fidelity} &
\textbf{Dynamicity} &
\textbf{Deployability} \\
\hline
Simplified analytical / datasheet-based \cite{Ferrari2020GNPy,bromage2004raman} &
Limited spectral and loading-dependent detail; suitable for nominal conditions. &
Fast steady-state evaluation; limited adaptation without recalibration. &
High with limited data and parameter requirements. \\
\hline
Calibrated parametric \cite{Borraccini2025differential} &
Good accuracy within characterized operating ranges; dependent on calibration coverage. &
Fast inference; periodic or event-driven parameter updates. &
High when sufficient spectral measurements are available. \\
\hline
Detailed physics-based \cite{Giles1991Modeling,bromage2004raman} &
Detailed physical representation within validated assumptions, including EDFA dynamics and distributed Raman effects. &
Can predict changing states; numerical and synchronization costs vary. &
Limited by availability of device, pump, and fiber parameters. \\
\hline
Data-driven \cite{wang2023open,zibar2019machine,Wang2025cascaded} &
Potentially high within the training distribution; uncertain extrapolation and cascade accuracy. &
Fast inference; adaptation may require additional data or retraining. &
Moderate; depends on characterization effort and transferability. \\
\hline
Hybrid / physics-informed \cite{Liu2023GrayBox} &
Combines physical consistency with measured behavior; depends on model structure and calibration. &
Supports adaptive predictions when telemetry and updates are sufficiently frequent. &
Moderate to high; depends on observability and calibration effort. \\
\hline
\end{tabular}
\end{table*}

Optical amplifiers are key components of wavelength-division-multiplexed (WDM) transmission systems, determining the generation of signal power and accumulation of ASE noise along the lightpath while compensating for optical losses.
Therefore, accurate amplifier modeling is critical for ONDT since gain and noise figure (NF) directly affect OSNR, nonlinear propagation, and QoT estimation.
Erbium-doped fiber amplifiers (EDFA) and Raman amplifiers represent the two main amplification techniques in optical transmission networks.
Although their physical mechanisms are different, both present a common modeling challenge to reproduce wavelength-dependent amplification under various operating conditions while balancing accuracy, computational complexity, and characterization requirements.
EDFAs provide lumped amplification through stimulated emission in erbium-doped fibers. 
Their gain and NF characteristics depend on population inversion, input power, spectral loading, and operating settings. 
Commercial devices also incorporate gain-flattening filters, variable optical attenuators, and automatic control mechanisms, and their internal parameters are often proprietary or inaccessible.
As a result, their observable behavior can differ from simplified physical descriptions, especially due to gain saturation, residual spectral ripple, and spectral hole burning.
Accurate device-level predictions alone cannot guarantee E2E accuracy, as these load-dependent effects can accumulate across cascaded amplifiers ~\cite{Wang2025cascaded}.

Conversely, Raman amplification relies on SRS to transfer optical power from the high-frequency pump to the low-frequency signal.
Unlike lumped EDFAs, distributed Raman amplifiers provide gain along the transmission fiber, intrinsically coupling amplification with longitudinal signal power evolution.
Their operation depends on pump wavelength and power, propagation direction, fiber properties, and WDM load.
Pump depletion and pump-pump interaction further affect the gain profile, especially in broadband and multiband systems.
Therefore, Raman amplifier modeling is closely related to fiber propagation modeling, which affects both OSNR and NLI estimation ~\cite{bromage2004raman}.
Physics-based models explicitly describe the mechanisms governing optical amplification.
For EDFAs, rate equation models reproduce wavelength-dependent gain, population inversion, and saturation effects, while dynamic expansion can account for response to channel add/drop events.
For Raman amplifiers, coupled propagation equations describe the spatial evolution of pump and signal power, including their interaction and distributed gain generation~\cite{bromage2004raman}. 
For ONDTs, the main difference is whether the required internal EDFA parameters or longitudinal Raman coefficients can be identified from accessible measurements.

Simplified analytical and parametric models reduce this complexity by representing amplifier operation through effective gain, NF, and spectral response functions.
For EDFAs, these abstractions may depend on vendor specifications or measured gain and noise profiles, complemented by calibration parameters that account for operational gain, slope, and spectral loading.
Similar simplifications apply to Raman amplification through approximate gain expressions and reduced propagation models.
These representations support repeated network-level QoT evaluations~\cite{Ferrari2020GNPy,Borraccini2025differential}.
Data-driven models learn relationships between operating conditions and resulting gain or noise characteristics from measured or simulated datasets.
Machine-learning techniques, including neural networks, have demonstrated accurate prediction of EDFA gain spectra and Raman amplification profiles with low inference complexity after training~\cite{wang2023open,zibar2019machine}.
They can represent devices with inaccessible internal parameters, but amplifier cascade errors remain an ONDT-specific concern~\cite{Wang2025cascaded}.
Hybrid and physics-informed approaches combine physical constraints with measurement-driven calibration. 
Rather than learning the entire amplifier response, these methods may identify effective physical parameters, incorporate analytical relationships into learning, or estimate residual corrections to an existing model. 
For EDFA gain prediction, calibrated gray-box formulations have been demonstrated~\cite{Liu2023GrayBox}; for Raman models, a corresponding physical baseline can use coupled pump--signal propagation equations. 
The two cases differ in what internal states can be inferred from operational telemetry.

From an ONDT perspective, the suitability of these approaches can be assessed through the FDD framework.
Importantly, these dimensions must be interpreted in relation to the intended ONDT function and the observability of the underlying physical system.
Fidelity refers to the ability to reproduce amplifier characteristics relevant to the target application. 
For EDFAs, accurate gain and NF spectra are essential for estimating per-channel power and OSNR, particularly under changing spectral loading and across cascades. 
For distributed Raman amplifiers, longitudinal power evolution is an additional requirement: different gain distributions may produce similar output powers while inducing different nonlinear propagation penalties. 
Thus, the appropriate physical detail depends on whether the ONDT supports component characterization, E2E QoT estimation, or transmission-system optimization. 
A model accurately predicting output spectra may suffice for some network-level decisions yet be inadequate when internal power distribution or transient response must be reproduced.
Dynamicity concerns the model's ability to remain synchronized with the physical system as operating conditions evolve. 
It includes both the time required to evaluate a new operating state and the ability to acquire sufficiently recent measurements, update parameters, and represent relevant physical dynamics. 
In EDFAs, channel add/drop events can produce fast gain transients followed by control-loop stabilization, whereas Raman amplification responds to changes in pump settings and spectral loading through coupled propagation and control mechanisms~\cite{Bononi1998reservoir}.
For quasi-static planning, rapid steady-state predictions may suffice. 
In contrast, real-time power optimization, reconfiguration, or transient analysis may require more frequent synchronization and sometimes explicitly time-dependent models. 
Fast inference alone therefore does not imply high Dynamicity if the model relies on outdated measurements or expensive recalibration.
Deployability describes the feasibility of integrating, calibrating, and maintaining a model within an operational optical network. 
Observability also distinguishes the two technologies: for commercial EDFAs, internal pump states and control parameters may remain hidden, favoring externally calibrated input--output representations; for distributed Raman amplifiers, pump settings may be accessible, but uncertainty in fiber properties and longitudinal power evolution still constrains physical modeling.

These considerations show that model selection is not determined solely by amplifier technology or intrinsic accuracy. 
It also depends on the information available from the physical infrastructure and on the operational decisions supported by the ONDT. 
For example, a calibrated input--output model may offer sufficient Fidelity and strong Deployability for steady-state QoT estimation, whereas rapid spectral reconfiguration requires loading-aware models and suitable synchronization mechanisms.
Likewise, nonlinear-performance optimization with distributed Raman amplification may require physically consistent longitudinal power profiles even when a simpler model reproduces the measured output gain.
Table~\ref{tab:optical_amplifier_fdd} summarizes the principal FDD trade-offs. The assessments are qualitative and depend on the application, available telemetry, and validated operating domain; they should not be interpreted as universal rankings.

\subsubsection{WSS Modeling}
\label{WSS Modeling}

WSSs are key components of ROADM, enabling lighthpath switching without optical-electrical-optical conversion at intermediate nodes \cite{Marom_Neilson_Greywall_2005,Strasser2010WSS}. 
They are therefore central to fixed-grid WDM, flex-grid, and superchannel-based optical networks, where lightpaths may traverse several ROADM nodes before reaching the receiver.
From a technological perspective, most WSSs can be interpreted as disperse-switch-recombine optical subsystems. 
The input WDM signal is collimated and spectrally dispersed by a diffraction grating or grism, so that different optical frequencies are mapped onto different spatial positions. 
A programmable switching element then redirects each spectral slice toward the selected output port, possibly imposing a controlled attenuation or shaping the spectral aperture, and the optical signal is finally recombined and coupled back into the output fiber. 
Early WSS implementations were mainly based on free-space optics and micro-electro-mechanical-system (MEMS) micromirror arrays, whereas modern flex-grid WSSs widely exploit liquid-crystal-on-silicon (LCoS) spatial light modulators, whose programmable phase patterns enable fine spectral granularity, variable slot widths, and adaptive passband shaping \cite{Marom_Neilson_Greywall_2005,Strasser2010WSS,ma_recent_2020,Robertson_Yang_Redmond_2014}. 
In ROADM nodes, WSSs are combined with optical amplifiers, monitoring taps, splitters/combiners, making their behavior strongly dependent on the selected route, output port, frequency slot, and attenuation setting \cite{Strasser2010WSS,Yang_Robertson_Wilkinson_Chu_2017}.

These same enabling technologies are also the origin of the main WSS-related physical impairments that must be represented in an ONDT. 
The WSS introduce insertion loss, and is not an ideal rectangular optical filter, having a frequency-dependent transfer function with finite bandwidth, possible center-frequency offsets, and passband asymmetry \cite{Tibuleac_Filer_2010,Pulikkaseril_2011_switches,Pulikkaseril_2011_networks}. 
When several ROADMs are crossed, the corresponding WSS transfer functions are cascaded, leading to passband narrowing, edge clipping, filtering penalties, and possible in-band or adjacent-channel crosstalk \cite{Tibuleac_Filer_2010,Hsueh_Stark_Liu_Tibuleac_Filer_Chang_Ralph_2012,Filer_Tibuleac_2012,heismann_system_2010,pilori_real-time_2020}.
Although WSSs are designed to operate as polarization-insensitive components, they may exhibit residual polarization-dependent transmission. 
This aspect is particularly relevant for LCoS-based WSSs, because the liquid-crystal phase modulation mechanism is intrinsically polarization dependent, contributing to PDL \cite{Strasser2010WSS,Zhu_Wen_Li_Chen_Yu_2023}.
Accordingly, WSS modeling in an ONDT should not treat the device as an ideal wavelength router. 
The WSS DT should represent at least: a scalar or configuration-dependent insertion loss for power and OSNR estimation; a port-, frequency-, and route-dependent PDL description for polarization-induced power and SNR variability; and a frequency-domain passband model for filtering penalties, cascaded-WSS narrowing, and residual distortions after coherent-receiver equalization. 

\begin{table*}[t]
\centering
\caption{WSS modeling choices along the FDD framework.}
\label{tab:wss_fdd}
\label{tab:wss_pdl_fdd}
\label{tab:wss_filtering_fdd}
\scriptsize
\setlength{\tabcolsep}{3pt}
\renewcommand{\arraystretch}{1.0}
\begin{tabular}{
|>{\raggedright\arraybackslash}p{\dimexpr0.085\textwidth-2\tabcolsep-1.2\arrayrulewidth\relax}
|>{\raggedright\arraybackslash}p{\dimexpr0.215\textwidth-2\tabcolsep-1.2\arrayrulewidth\relax}
|>{\raggedright\arraybackslash}p{\dimexpr0.25\textwidth-2\tabcolsep-1.2\arrayrulewidth\relax}
|>{\raggedright\arraybackslash}p{\dimexpr0.22\textwidth-2\tabcolsep-1.2\arrayrulewidth\relax}
|>{\raggedright\arraybackslash}p{\dimexpr0.23\textwidth-2\tabcolsep-1.2\arrayrulewidth\relax}|}
\hline
\textbf{Family} &
\textbf{Model} &
\textbf{Fidelity} &
\textbf{Dynamicity} &
\textbf{Deployability} \\
\hline

\textbf{PDL} &
Worst-case PDL margin &
Low; conservative worst-case accumulation &
Route/device changes &
Very high; datasheet PDL and deterministic accumulation \\
\cline{2-5}

&
SNR probability density function with worst-case device PDL \cite{Miotto2026ExactPDLSNR} &
Medium; random state of polarization (SOP) and distributed noise; fixed device PDL &
Route/noise-source configuration changes &
Medium; worst-case PDL, noise-source data, analytical PDF \\
\cline{2-5}

&
Moment-matched SNR with Maxwellian device PDL \cite{Miotto2026Disaggregable} &
High for port/frequency variability; approximate SNR statistics &
Route, port/slot, or PDL-statistics changes &
Medium; characterized PDL statistics and moment matching \\
\hline

\textbf{Spectral response} &
Super-Gaussian passband \cite{torres-ferrera_impact_2025} &
Low--medium; nominal bandwidth/roll-off; limited device detail &
High with available center/bandwidth parameters &
Very high; three-parameter model \\
\cline{2-5}

&
Aperture response (error-function model) \cite{miotto_generalized_2026} &
Medium--high; device-aware passband and rounded edges &
High with programmed center/bandwidth and calibrated response &
High; analytical model calibrated from optical spectrum analyzer (OSA) traces \\
\cline{2-5}

&
Measured amplitude response &
High for characterized passbands and device-specific features &
Limited by measurement coverage and refresh &
Medium--low; characterization, storage, possible vendor-specific data \\
\hline

\textbf{Data-driven} &
Learned WSS/ROADM surrogate \cite{Sequeira2023OCATA} &
Potentially high within training/calibration domain &
Fast inference; new configurations may require new training data &
Medium; training data, calibration, extrapolation control \\
\hline
\end{tabular}
\end{table*}


Starting from the power dimension identified above, the lowest-complexity WSS model represents the device through a scalar insertion-loss value.
This value can be obtained with optical power meters or through OCMs placed before and after the WSS. In an ONDT, this loss term updates the power evolution along the lightpath and contributes to OSNR estimation.
As anticipated, WSSs may also introduce PDL, namely a differential attenuation between orthogonal states of polarization of the transmitted signal. In optical transport systems, PDL may originate from several optical components, but WSSs inside ROADMs are particularly relevant because they are repeatedly crossed by express and add/drop channels in reconfigurable networks \cite{Strasser2010WSS,DAmico2022WSSPDL,Borraccini2023WSSPDL}.
The instantaneous PDL penalty depends on the state-of-polarization (SOP) entering the device, while the SOP itself evolves randomly along the fiber because of birefringence, mechanical stress, and temperature variations. Moreover, experimentally characterized ROADM PDL can depend on the selected port and optical frequency, so that a WSS PDL model should be associated with the specific routing state, output port, and spectral slot of the lightpath \cite{DAmico2022WSSPDL,Borraccini2023WSSPDL}.
The representation of PDL concentrated at component level is the so-called "hinge model" for the PDL \cite{Zarkovsky2020HingePDL}. The key idea of such mathematical abstraction is to represent the attenuation experienced by one recovered polarization tributary as a random variable. The statistics of the attenuation depends on experimentally characterized parameter, commonly referred to as $\mathrm{PDL}_{\mathrm{dB}}$, whose worst case value can be found on the element datasheet. 
This model is compact enough for system-level ONDT implementations, and allows to analytically express the PDL-induced SNR distribution \cite{Miotto2026ExactPDLSNR}. As a further step, instead of using the worst case datasheet value of $\mathrm{PDL}_{\mathrm{dB}}$, it is possible to account for its variability due to port/frequency combination\cite{Borraccini2023WSSPDL}. Recent studies have modeled the $\mathrm{PDL}_{\mathrm{dB}}$ as a Maxwellian random variable, approximating the PDL-induced SNR distribution with moment matching techniques\cite{Miotto2026Disaggregable}.
At link level, PDL affects both the useful signal and the ASE noise contributions accumulated before and after each PDL element\cite{DAmico2023PDL_OSNR}. Therefore, PDL is observed as a stochastic fluctuation of the received power, OSNR, and post-equalization SNR\cite{Nelson2011PDLStatistics}. Traditional engineering practice can include a worst-case PDL margin based on the accumulated PDL along the path, but this approach is often unnecessarily conservative because of the low probability of extreme PDL events. A more informative ONDT model propagates the PDL-induced SNR distribution and derives the operating margin from a target probability of out of service (POOS) \cite{Shtaif2008CoherentPDL,Miotto2025PDLMargin}.
Crucially, an accurate \emph{PDL penalty} is not determined by the PDL statistics alone, due to equalization action on the PDL-affected signal and noise. Consequently, a high-fidelity ONDT should condition the PDL-induced QoT penalty on the receiver noise level and on the equalization strategy (see Sec. \ref{TRx Modeling}).
Several levels of PDL modeling can be adopted depending on the target ONDT function. A low-complexity model uses the stored per-device worst case $\mathrm{PDL}_{\mathrm{dB}}$ values to compute a conservative deterministic margin for fast control but limited fidelity. An intermediate model evaluates the PDL-induced SNR probability density function from the cascade of PDL elements and distributed noise sources, assuming the worst case per-device $\mathrm{PDL}_{\mathrm{dB}}$\cite{Miotto2026ExactPDLSNR}. 
A high-fidelity model can account for the frequency/port variability, modeling the $\mathrm{PDL}_{\mathrm{dB}}$ as a random variable and approximating the PDL-induced SNR distribution through moment matching techniques\cite{Miotto2026Disaggregable}.


The second modeling dimension concerns the frequency-selective response of the WSS.
First-order models describe WSS frequency selectivity through a linear filter’s amplitude response, obtained from optical spectrum analyzer (OSA) measurements or parametric fits~\cite{pilori_real-time_2020}. Super-Gaussian models use center frequency, bandwidth, and order, offering compact approximations for QoT and cascaded-WSS analyses but potentially missing real device characteristics~\cite{maniloff_3cw_2023_spectral_isolation,heismann_system_2010}. Error-function models convolve the programmed spectral aperture with the WSS optical response and use center frequency, bandwidth, and shape factor, typically achieving closer agreement with measurements~\cite{Pulikkaseril_2011_switches,Pulikkaseril_2011_networks}. Directly measured responses provide the highest fidelity but require substantial storage and potentially vendor-specific data. The WSS transfer function describes the device, whereas the filtering penalty depends on the complete impairment--receiver chain, as discussed in Section~\ref{TRx Modeling}.

When assessing the FDD principles on WSS modeling dealing with ONDT development, it is pivotal to emphasize the interplay of different phenomena.
Fidelity increases when the ONDT moves from single worst-case PDL margin to port-, frequency-, and route-aware statistical SNR modeling. The same can be said when including in the filtering penalty modeling erfc-based filtering models, TRx impairments and DSP effects.
Dynamicity requires the PDL model to be updated when the route, WSS port, center frequency, slot width, or device configuration changes, and also requires the ONDT to account for SOP-induced QoT fluctuations even when the lightpath itself remains unchanged. In the case of passband, the model should track WSS reconfigurations, including changes in center frequency, output port, slot width, attenuation setting, and filtering profile.
From a deployability perspective, practical ONDTs should support: deterministic PDL margins for robust brownfield deployment, statistical approximations using worst-case device PDL for scalable network optimization, and SNR distributions capturing port/frequency variability for low-margin operation and detailed QoT validation. For passbands, error-function models balance Fidelity and Deployability, while measured responses offer the highest accuracy but require extensive characterization, greater storage, and vendor-specific information.


A complementary route is to learn the WSS/ROADM behavior from data rather than to parameterize every impairment explicitly. 
OCATA models optical links and ROADMs by concatenating pretrained neural component models with measured WSS responses \cite{Sequeira2023OCATA}. It lacks a general interpretable filtering-penalty formula; extrapolation is uncertain, and hardware/configuration changes may require new data. Hybrid ONDTs can combine physics-based passband/PDL models for extrapolation and disaggregation with learned residual or node-level models for proprietary effects.

\subsection{Monitoring}
\label{Monitoring}
Section~\ref{Modeling} described how the optical physical layer is expected to behave according to a-priori models. A real network, however, rarely matches such a description even at the time of deployment, because span losses, fiber types, connector and splice losses, and amplifier operating points often differ from their design values or inventory records, and some parameters are vendor-proprietary and not disclosed~\cite{Sasai2025ONTMultiDomain, Ferrari2020GNPy}. The network then continues to depart from it over time through aging, reconfiguration, temperature and polarization fluctuations that act over widely different time scales. Monitoring provides the ONDT with measured ground truth against both and calibrates the models to the network as actually built (Fidelity) and keeps them synchronized with the network as it changes (Dynamicity), and it forms the observe stage of the control loop discussed in Section~\ref{Network Control and Actuation}. In optical networks,
this physical state is distributed over hundreds of kilometers of largely
inaccessible infrastructure. From the viewpoint of the ONDT, monitoring techniques should therefore be assessed in terms of the physical quantities they provide as inputs to the models in Section~\ref{Modeling}, the spatial coverage, spatial granularity, and update interval of those quantities, and the cost of their deployment in an operational network. In this subsection, we first briefly review conventional monitoring techniques from this viewpoint and then focus on DLM, which makes a transmission link observable from the TRx. Table~\ref{tab:monitoring_fdd_M_A} summarizes these techniques along the FDD framework.

\begin{table*}[htbp]
\centering
\caption{Qualitative FDD assessment of monitoring techniques as sources of input to the ONDT.}
\label{tab:monitoring_fdd_M_A}
\scriptsize
\setlength{\tabcolsep}{3pt}
\renewcommand{\arraystretch}{1.26}
\begin{tabularx}{\textwidth}{|
>{\raggedright\arraybackslash}p{2.1cm}|
>{\raggedright\arraybackslash}X|
>{\raggedright\arraybackslash}X|
>{\raggedright\arraybackslash}p{2.6cm}|
>{\raggedright\arraybackslash}p{1.8cm}|
>{\raggedright\arraybackslash}X|}
\hline
\multirow{2}{*}{\textbf{Technique}} & \multirow{2}{*}{\shortstack[l]{\textbf{Quantities provided}\\\textbf{to the ONDT}}} & \multirow{2}{*}{\textbf{Fidelity}} & \multirow{2}{*}{\textbf{Dynamicity}} & \multicolumn{2}{c|}{\textbf{Deployability}} \\
\cline{5-6}
 & & & & \textbf{Maturity} & \textbf{Attachability} \\
\hline
DSP-based TRx telemetry \cite{Hauske2009OPM,OIF2025CCMIS,OpenConfig} &
Link-accumulated CD, PMD, PDL, SNR, and pre-FEC BER; end-to-end constraints or references for QoT estimates. &
High for end-to-end values; no attribution to individual spans or amplifiers. &
High; continuous measurements can be streamed through standardized telemetry. &
Very high; widely deployed in coherent TRxs. &
High; available from deployed coherent TRxs through supported management interfaces. \\
\hline
OCM and amplifier photodiodes \cite{He2023liveQoT} &
Per-channel or total optical power at equipped nodes; amplifier gain and tilt. &
High at equipped nodes; no direct observation between monitoring points. &
Medium--high; update interval depends on line-system control and polling. &
Very high; established monitoring functions. &
High where equipped; coverage and data access depend on installed hardware and vendor interfaces. \\
\hline
OTDR (stand-alone or embedded) &
Distributed fiber loss profile; loss events and reflections within a span. &
High within a span, with meter-scale resolution; observes loss at the probe wavelength rather than data-signal power. &
Medium; continuous when supported by embedded monitoring, otherwise on demand. &
Very high; established field measurement technique. &
Low--medium for unequipped spans; link-wide coverage generally requires instruments at individual spans (with submarine HLLB exceptions). \\
\hline
DLM \cite{Sasai2022DLM,Sasai2026PluggableDLM,Pacini2025SelfHealing} &
Distributed signal-power profile along the link; span loss, amplifier gain, anomaly location; extensions to PDL, DGD, and Raman gain. &
Medium--high; km-scale resolution; accuracy depends on launch power, SNR, and knowledge of fiber nonlinear coefficients. &
Medium--high; averaging-time/accuracy trade-off; 1.5-s profile streaming demonstrated. &
Emerging operational integration; demonstrated in live networks and commercial coherent DSPs. &
Potentially high with supported TRxs; in-service monitoring without added intermediate hardware, including across administrative boundaries; telemetry model not yet standardized. \\
\hline
\end{tabularx}
\end{table*}

DSP-based TRx telemetry yields link-accumulated quantities such as CD, PMD, PDL, SNR, and pre-FEC BER~\cite{Hauske2009OPM}, which are exposed through standardized management interfaces and telemetry models such as Coherent CMIS~\cite{OIF2025CCMIS} and OpenConfig~\cite{OpenConfig}. These quantities enter the ONDT as E2E constraints on the models in Section~\ref{Modeling}, or as a reference against which the QoT estimated by the ONDT is verified. 
Being accumulated over the whole link, these quantities provide E2E information for QoT estimation and validation, while per-element calibration requires additional spatially resolved information.
OCMs and photodiodes of amplifiers provide per-channel or total power in fibers at each node. They are the principal input for calibrating amplifier gain and tilt, and they have enabled production-scale refinement of QoT estimators (Sec.~\ref{Validation}, \cite{He2023liveQoT}). 
Their granularity corresponds to the nodes at which they are installed, providing node-level information for model calibration.
OTDR, either as a stand-alone instrument or as a function embedded in the line system, yields a distributed fiber loss profile and locates loss events and reflections within a fiber span at meter-scale resolution.
Link-wide coverage requires an instrument at every span, which may affect Deployability, particularly in legacy systems (except in submarine systems with high-loss loopback configuration).
Moreover, the measurement is confined to the distributed and lumped losses of the fiber at the probe wavelength, and does not observe the power of the data signal itself. 
These techniques provide complementary monitoring capabilities at different spatial scales.
DLM further extends these capabilities by estimating the physical state distributed along the link simply by processing the signal received by a coherent TRx at its end, as introduced in section~\ref{A New Monitoring Paradigm: The Transceiver as a Distributed Sensor}

Its spatial resolution is enabled by CD and the fiber Kerr nonlinearity~\cite{Sasai2022DLM,Sasai2023PPElimit,Tanimura2020AnomalyID}. Because of CD, the signal waveform varies along the fiber, and the Kerr nonlinearity generates a nonlinear waveform from this local waveform in proportion to the local signal power. The nonlinear waveform from each position thus carries a distinct signature, and by separating the received nonlinear waveform into these contributions in the DSP, the power at each position, i.e., distributed power profile along the fiber, can be estimated. Since the computation is performed in the receiver DSP on the data-carrying signal itself, DLM requires neither probe light nor instruments at intermediate sites, and it operates in service.

The power profile obtained by DLM can be supplied to the models in Section~\ref{Modeling} in several forms, depending on the input that each model takes: the profile itself, the loss coefficient or total loss of each span, the launch power into each span, or the gain of each amplifier~\cite{Sasai2022DLM,Sasai2024LLS}. These values calibrate the span parameters of the fiber model and the operating point of the amplifier model (Sections~\ref{Fiber Propagation Modeling} and~\ref{Optical Amplifier Modeling}), which is a prerequisite for accurate QoT estimation by the GN model~\cite{Curri2022GNPyPhysicalLayer,Borraccini2024EDFAcalib}. In a field trial, calibrating the line system with DLM-derived values reduced the QoT estimation error from about $\pm$0.6 to $\pm$0.3~dB~\cite{Nishizawa2024semi-automatic}.  Because the profile is resolved in distance, link anomalies, such as excess connector or splice loss, fiber bending, or amplifier gain deviation, can also be identified together with the position at which it occurs~\cite{Tanimura2020AnomalyID}, so that a discrepancy between the model and the measurement can be traced to a physical cause rather than absorbed into a residual term~\cite{Pacini2025SelfHealing}. This also elevates fault management from the detection of symptoms at the endpoint to the direct isolation of the causal location, enabling fault isolation within an otherwise inaccessible transmission line without additional equipment \cite{Pacini2025SelfHealing}. The same DLM principle has been extended to other distributed quantities that enter the models: the value and location of PDL~\cite{Eto2022PDLloc,Takahashi2024PDL,Andrenacci2023PDL}, DGD~\cite{Hahn2024DGD}, the longitudinal Raman gain profile discussed in Section~\ref{Optical Amplifier Modeling}~\cite{Kaneko2025Raman}, and the joint dependence on distance, time, frequency, and polarization~\cite{Sasai2025Tomography4D}. Its applicability has also been confirmed over links composed of several fiber types~\cite{Takahashi2025VariousFiber,Pilori2025MixedFiber}.

In terms of maturity, DLM has been demonstrated over a live production network spanning several administrative domains~\cite{Sasai2025ONTMultiDomain}, and in 2026 it was implemented in a commercial coherent DSP and pluggable TRx, providing km-scale resolution over a link of about 1,000~km as a native function of the TRx~\cite{Sasai2026PluggableDLM}. At the research level, the measurement range has been extended beyond 10,000~km for submarine links~\cite{May2023Submarine10000km}, and a resolution of 20--40~m has been reported over a 302-km link~\cite{Sasai2025TensOfMetre}. Because DLM runs in the DSP of pluggable TRxs, it can be applied to various networks from short-reach DCI to long-haul links without additional hardware~\cite{Sasai2026PluggableDLM,Nishizawa2026JOCN}, and because it relies only on the operator's own endpoint, it can observe segments beyond an administrative boundary where instruments at intermediate sites are not accessible~\cite{Sasai2025ONTMultiDomain}. Field trials in which DLM is integrated into an ONDT, for line-system provisioning, hybrid physics/ML QoT prediction, and multi-domain observability, are reviewed in Section~\ref{Validation}.

From the FDD perspective, the Fidelity of DLM depends on the link condition: the estimation relies on the nonlinear component of the received signal, so its accuracy decreases at low launch power and low SNR~\cite{Sasai2023PPElimit}. In addition, what is measured is strictly the product of the fiber nonlinear coefficient and the local optical power, so prior knowledge of the nonlinear coefficient of each fiber is required when the absolute power is of interest. Within a single span, its resolution and dynamic range do not reach those of OTDR. In practice, the spatial resolution is on the order of 1--3~km for TRxs operating at around 120~GBd, and the commercial implementation provides a sampling resolution of 2.1~km over a 1005-km link, and the estimated profiles agree with OTDR measurements within an RMS error of 0.4--0.6~dB\cite{Sasai2026PluggableDLM}. In terms of Dynamicity, the update interval is set by the number of samples averaged to suppress noise, and there is a trade-off between averaging time and accuracy. 
Streaming of power profiles from the coherent receiver to the SDN controller every 1.5~s has been demonstrated, together with its use for automatic fault localization and correction~\cite{Pacini2025SelfHealing}, which brings DLM within the time scale of the control loop discussed in Section~\ref{Network Control and Actuation}. Its Deployability is high because the measurement is made in service on the data signal and already implemented on a coherent DSP of pluggable transceivers~\cite{Sasai2026PluggableDLM}. It can be repeated continuously without a maintenance window and additional hardware. DLM therefore complements rather than replaces the techniques described earlier, and an ONDT can use OCM and OTDR data where they exist and fill the remaining segments with DLM, and the resulting distributed state is passed to the control loop in Section~\ref{Network Control and Actuation}, where it has begun to be used for improved optimization and resilience of optical networks~\cite{Mano2026Resilience}.


\subsection{Data Management, Network Control and Actuation}
\label{Network Control and Actuation}

The ONDT control loop can be described as an observe--estimate--decide--act process. Monitoring and telemetry collect the operational state from various network elements. The ONDT estimates hidden physical quantities, calibrates model parameters, predicts QoT, and evaluates candidate actions. The network control and actuation function then enforces the selected action. In doing so, parameter/state uncertainty caused by synchronization issues or missing telemetry data could introduce an error in ONDT. This type of error is different from model-error stemming from analytical model approximations. These are two major sources of ONDT estimation errors.


The recent research activities in the data management, network control and actuation field focus on open challenges including network discovery and multi-entity control~\cite{Kani2026OpenROADM}, governance of data and control in multi-stakeholder ecosystems~\cite{zaid2026ecoc,Mitrovska2026Ownership}, as well as keeping the ONDT constantly updated with real-time telemetry in a reliable way~\cite{Shen2025UnifiedTelemetry}. The authors in~\cite{Kani2026OpenROADM} describe user-site transponders and their remote control, which is suitable for IPoWDM deployments in multi-entity settings and helps keep the digital representation aligned with the physical network. An alternative way is presented in~\cite{Mitrovska2025Governance}, in which  cross-stakeholder control sharing, access rights, and administration boundaries are modeled as enforceable policies to establish trustworthy and compliant data use. These aspects are critical for deployability of ONDT in multi-stakeholder environments.


Therefore, a major research challenge is not only to design higher-performing SDN controllers, but also to design ONDT control loops that are secure, privacy-preserving, telemetry-consistent, and vendor-interoperable. Such loops should optimize lightpaths without requiring operators to disclose sensitive topology or inventory data or vendors to expose their own TRx and OLS internals. Promising directions include abstract but physically meaningful data models, federated or privacy-preserving QoT estimation, secure telemetry pipelines with authentication and freshness guarantees, controller-side reliability scoring, validation of AI-generated actions before launch, and standardized mechanisms for exposing only the minimal physical parameters required by ONDT.
In this sense, SDN control is the necessary operational substrate for ONDT, while the twin provides the physical-layer intelligence required to make SDN decisions reliable in open, disaggregated, and low-margin optical networks.


\subsection{AI Agent} \label{ai_agent}

With the advancement of Agentic AI, observability is no longer a static property of the monitoring system but rather depends on action. Digital twins provide the basis for predicting the outcome of a proposed action before executing it on the actual network. This enables virtual experimentation and safe closed-loop control.
This section evaluates AI-agent-related technologies for optical network operation with FDD framework, focusing on their capabilities, limitations, and practical applicability.

%
Optical-network tasks like performance prediction and fault diagnosis are usually based on individual models for each task and each dataset.
Foundational models provide a single, unified representation of the disparate knowledge contained within a network.
Retrieval-augmented generation (RAG) connects these models to external knowledge sources, enabling them to retrieve relevant network records, equipment specifications, and operational procedures when answering queries or supporting decisions.
The key for the effectiveness of foundation models is the integration of different types of information, which are usually considered in isolation.
%
Reasoning models build on this capability by performing multi-step diagnosis, planning, and cross-layer inference.
It can help connect network observations to causes and actions for corrective measures. They rely on having access to accurate network information and on systematic validation of conclusions reached. Such validation is typically conducted using tools and simulators. Models that are able to request additional measurements, evaluate alternative actions and update their plans in the light of the results of these evaluations are reasoning-capable models integrated into an agentic loop.
%
AI agents represent a much more fundamental shift to act on the network by performing tasks such as probing, selecting measurements, changing routes, and modifying configurations.
It extends network analysis through a feedback loop where they select actions, observe their effects, and use the resulting information to guide subsequent decisions. For example, in an optical network, an agent can request telemetry, investigate faults, change the measurement parameters, or change the device parameters. These actions help to differentiate between different hypotheses when the current measurements are insufficient. Also, routing and configuration changes can modify the network state and thus the conditions for subsequent measurements.

\begin{table*}[t]
\centering
\renewcommand{\arraystretch}{1.3}
\caption{Comparison of LLM-Based and Agentic AI Approaches for Optical Networks: Adaptation Approach and FDD Characteristics}
\label{tab:fdd_comparison}
\scriptsize
\resizebox{\textwidth}{!}{%
\begin{tabular}{|>{\raggedright\arraybackslash}p{3.0cm}|>{\raggedright\arraybackslash}p{3.6cm}|>{\raggedright\arraybackslash}p{3.2cm}|>{\raggedright\arraybackslash}p{3.2cm}|>{\raggedright\arraybackslash}p{3.2cm}|}
\hline
\textbf{Agent Approach} & \textbf{Adaptation Approach} & \textbf{Fidelity} & \textbf{Dynamicity} & \textbf{Deployability} \\
\hline
Text-only LLM assistant~\cite{wang2024alarmgpt,pang2024llama2loganalysis} &
Prompting + RAG, or
Local instruction tuning &
no DT; text domain (alarm/log) only &
no DT; no physical-layer synchronization &
Medium; carrier data with single-GPU deployment \\
\hline
Simulation-coupled agent~\cite{zhang2025llmagent,zhou2024sdnagent,Tanimura2024LLM} &
Prompting optimization + external function calling &
Medium; GN-model emulation without hardware &
Low; simulated telemetry only, no live re-sync &
Low; simulation-only \\
\hline
DT-free hardware-control agent~\cite{wang2026agentic} &
Prompting + Model Context Protocol based tool calling &
no DT; directly controls physical instruments &
High; continuous live hardware feedback &
High; real multi-vendor hardware \\
\hline
DT hardware-control agent~\cite{qiu2026expertiseguided,liu2025first,liu2026field,Song2025_LLMDT} &
Prompting + ReAct + RAG + DT&
High; DT calibration for network control &
High; DT re-synchronized from live telemetry &
Medium; field-deployed links, commercial hardware \\
\hline
\end{tabular}%
}
\end{table*}
The existing LLM-based and agent-based approaches for optical networks are categorized into four types (see Table~\ref{tab:fdd_comparison}). Text-only LLM assistants use prompts with RAG or local instruction tuning to analyze alarms and logs, but they do not utilize a DT or provide any hardware actuation~\cite{wang2024alarmgpt}. Simulation-coupled agents invoke QoT tools via external function calling or a fixed intent vocabulary, but have only been tested in simulations~\cite{zhang2025llmagent}. DT-free hardware-control agents operate directly on physical devices and obtain live feedback from the hardware rather than using a DT~\cite{wang2026agentic}. Hardware-validated DT-coupled agents employ a DT that is calibrated and re-synchronized based on network telemetry to assess actions on field-deployed networks, resulting in sub-dB QoT prediction errors~\cite{qiu2026expertiseguided, Song2025_LLMDT}.
In AgentOptics~\cite{wang2026agentic}, the feedback loop for controlling devices is implemented with optical hardware. An LLM is used to interpret the operational objectives of the network, to decompose these into a series of steps to be executed on heterogeneous devices, and to control the devices using Model Context Protocol (MCP) tools in the context of the model. The architecture is designed to integrate with existing control systems, to reliably execute the tools, and to provide several types of safeguards for the network.
An ONDT can support this architecture by estimating the network state and evaluating the candidate actions prior to hardware execution~\cite{Mano2024TRXROP,raj2024multi}. The agent can use these predictions to select an action, execute it using the available tools, and compare the measured outcome with the expected result. Any discrepancies inform further measurements or revisions to the plan, thereby closing the loop between prediction, intervention, and verification.
The changes an agent may make to a live optical network can have negative consequences beyond those of an incorrect diagnosis, i.e., they may interrupt traffic or degrade the QoT. Therefore, any proposed action needs to be evaluated against operational policies, device constraints and service requirements prior to its execution. Predictions of the effect of candidate actions on a network can be generated by world models. This would allow an agent to assess, for example, whether a proposed rerouting would maintain the QoT, prior to its implementation by the controller. These predictions are then used to select the best action, and controller validation and post-action measurements are required to account for possible model errors.

In summary, the ability of a self-experimental optical network to generate hypotheses, to perform constrained experiments, and to update its world model on the basis of measurements obtained during the experiments would enable it to function as a self-experimental network in the event of unexplained degradation.

\section{Validation}
\label{Validation}
The practical applicability of the enabling technologies discussed in the previous section must be evaluated through experimental and field validation.
The required functions and performance, as well as the challenges associated with ONDT deployment, vary significantly depending on the target application and operational environment.
This section organizes ONDT applications into two categories: “Enhancement of Conventional Optical Network Operations” and “Emerging Optical Network Services and Ecosystems.”
The former places importance on how ONDT can be integrated into a commercial optical network already equipped with advanced monitoring and control functions.
The latter is concerned with how flexible optical connections and new services can be achieved under limited information and observational ability in an environment that spans different users, providers, vendors, and/or administrative domains.
\subsection{Enhancement of Conventional Optical Network Operations}
\label{Enhancement of Conventional Optical Network Operations}

Accurate modeling of transmission paths in conventional optical networks requires consideration of multiple physical effects, such as power transfer between channels due to SRS (see \ref{Fiber Propagation Modeling}), wavelength-dependent gain of optical amplifiers and its accumulation (see \ref{Optical Amplifier Modeling}), and filtering by ROADM/WSS (see \ref{WSS Modeling}).
Commercial optical transmission systems have traditionally employed OCM and Optical Supervisory Channel (OSC) mechanisms, together with vendor-specific equipment parameters and control algorithms, to support transmission-path monitoring, optical-power adjustment, and optimization.
On deployment in such an environment, we can expect ONDT to combine the abundant amount of monitoring information obtained from the actual network with a physical model or data-driven model and estimate/update the physical parameters of optical amplifiers, fiber, etc.
In this way, ONDT can support transmission quality estimation, optical power optimization, and responses to faults and equipment aging while enhancing automation of the total network lifecycle.
%
At the same time, it is typical in such an application domain for the existing vendor-integrated system itself to be equipped with advanced monitoring, control, and optimization capabilities.
Thus, in moving toward actual ONDT deployment, it is important to evaluate ONDT not only in terms of model accuracy and optimization performance but also in what kind of added value it can provide to the existing commercial system and whether ONDT can coexist with existing facilities and operational workflows.

\begin{table*}[t!]
\centering
\scriptsize
\caption{
Representative Experimental and Field Validations of ONDT for Conventional Optical Network Operation}
\label{tab:ondt_validation_conventional}
\renewcommand{\arraystretch}{1.15}
\setlength{\tabcolsep}{3.5pt}
\begin{tabularx}{\textwidth}{|
    >{\raggedright\arraybackslash}p{1.20cm}|
    >{\raggedright\arraybackslash}p{1.55cm}|
    >{\raggedright\arraybackslash}p{2.05cm}|
    >{\raggedright\arraybackslash}p{1.95cm}|
    >{\raggedright\arraybackslash}p{3.20cm}|
    >{\raggedright\arraybackslash}p{1.05cm}|
    >{\raggedright\arraybackslash}X|}
\hline
\textbf{Study} &
\textbf{Validation environment} &
\textbf{Main ONDT functions validated} &
\textbf{Measurement / observability} &
\textbf{Main experimental results} &
\textbf{FDD viewpoint} &
\textbf{Significance and constraints in actual deployment} \\
\hline

SAMA-IR~\cite{Zhang2025SAMAIR} &
Field-deployed optical link &
Sensitivity-guided input refinement against many uncertain inputs &
Q-factor measurements; power measurements &
Improved Q-factor error from $-2.52\pm0.96$ to $-0.03\pm0.55$~dB using only 20 items of measurement data. &
\textbf{F/Dp} &
Improves Fidelity with a few measurements by prioritizing refinement targets according to their impact on QoT without having to uniformly refine all parameters. \\

\hline

Live-Net QoT Cal.~\cite{He2023liveQoT} &
Large-scale live production network: 76 OMSs, 694 EDFAs &
Production-scale calibration of analytical QoT estimator &
EDFA total power; OCM per-channel power; network telemetry &
GSNR estimation-error standard deviation: 0.60--0.63~dB baseline, 0.46~dB after input refinement, 0.24~dB after data-driven optimization &
\textbf{F/Dy/Dp} &
Shows the effectiveness of data-driven correction applied only to factors that are difficult to measure and to residual errors, while retaining a physics-based model; complex ML need not be uniformly introduced. \\

\hline

Field-Deployed DT~\cite{Song2024_FieldDeployedDT} &
Field-trial C+L-band optical transmission link &
Field-deployed DT implementation and operational recovery &
Field measurements; network/device parameters; status information &
Channel-power prediction accuracy improved by up to 2.3~dB and GSNR estimation accuracy by up to 1.3~dB; fiber-cut recovery demonstrated &
\textbf{F/Dy/Dp} &
Extends DT validation from offline accuracy to a stage that reflects network changes in the digital representation and feeds the results back to actual operations. \\

\hline

\multicolumn{7}{l}{
\textit{F: Fidelity; Dy: Dynamicity; Dp: Deployability.}
}
\end{tabularx}
\end{table*}

\subsubsection{Validation for Conventional Network Operation}

In an actual optical network, however, physical parameters are not always known or fixed, and the state of the network changes due to device-to-device variation, channel loading, monitoring error, vendor-specific behavior, environmental variation, aging, and other factors.
Thus, in validating the practical use of ONDT, it is important to evaluate not only the accuracy of the QoT model itself but also whether unknown or uncertain parameters can be estimated and updated from limited observational information and whether its digital representation can be synchronized with the physical network.
This deals with the issues described in the previous section, that is, when applying ONDT to an existing commercial network, it is necessary to consider not only model fidelity but also monitoring infrastructure, calibration procedure, and integration with existing operations. 
A modular DT combining QoT parameter refinement (input refinement; IR) and machine learning (ML) based EDFA modeling was experimentally tested on a WDM optical network~\cite{Morette2023MLDT}.
This approach enabled both current-state QoT estimation and network-state prediction after service addition/removal.
However, in large-scale optical networks, many uncertain parameters simultaneously exist across TRxs, fibers, and amplifiers, so uniformly and simultaneously refining all such parameters leads to an increase in required measurement data and optimization complexity.
Sensitivity-Analysis Guided Memory-Aware Input Refinement (SAMA-IR) was proposed and validated in field trials to prioritize stepwise refinement of uncertain parameters—including lumped losses, EDFA gain, tilt, NF, and TRx impairments—according to their impact on QoT~\cite{Zhang2025SAMAIR}.
As summarized in Table~\ref{tab:ondt_validation_conventional}, these results show that prioritizing refinement targets based on their impact
on system performance is effective in efficiently achieving high estimation
accuracy with limited measurement data.
%
Measurement-informed refinement was validated in a large live production network, demonstrating improved QoT estimation accuracy by combining measurement refinement and data-driven correction with a physics-based analytical model~\cite{He2023liveQoT} (Table~\ref{tab:ondt_validation_conventional}), it also showed that it is not necessarily appropriate to estimate all unknown parameters in a data-driven manner.
Although it investigated a method for learning EDFA NF from field GSNR, additional improvement to QoT prediction accuracy was nevertheless limited.
This result demonstrates the importance of maintaining a physics-based model as a foundation to work from while selectively applying data-driven correction to factors with insufficient observability.
%
%
In the above way, DT validation has evolved from field calibration and recovery at one point in time to lifecycle-wide synchronization that keeps up with changes in the physical network.
%
Field-deployed DT validation addresses uncertainty in parameter identification, device modeling, and state updates through measurement-informed refinement and DT-assisted reoptimization ~\cite{Song2024_FieldDeployedDT}, and extends to full-lifecycle synchronization, including device replacement~\cite{Song2025lifecycle}~(Table ~\ref{tab:ondt_validation_conventional}).

These studies demonstrate the evolution of ONDT validation from static model accuracy assessment to continuous synchronization with physical network changes.
Thus, in addition to initial model accuracy, ONDT Fidelity in actual operation must be evaluated on the extent to which the digital representation can be accurately and continuously synchronized with the physical network using limited observational information.
It should be noted here that the frequency and cost of measurement, calibration, and model updating necessary for this synchronization form direct trade-offs with Dynamicity and Deployability.

In conventional optical networks, vendor-integrated systems commonly provide operational functions such as monitoring, alarms, redundancy, configuration, and channel-power and QoT optimization (see~\ref{Control Architecture}).
As a result, the Deployability of ONDT in conventional network operation cannot be evaluated solely on whether physical-layer prediction and state synchronization can be executed in the actual network.
That is, among the functions handled by the existing system, it must be made clear as to which functions can be replaced, integrated, or enhanced by ONDT and which can be left as-is in the existing system and simply linked to ONDT.
Also important is whether ONDT can be deployed in a stepwise and realistic manner considering coexistence with existing facilities, vendor-specific functions, and existing operational workflows.
Because the implementation details and performance of monitoring, calibration, and optimization in vendor-integrated commercial systems are generally not publicly available, quantitative comparisons of ONDT's additional operational benefits with existing commercial solutions remain difficult.
It is therefore important in future validations to evaluate not only prediction accuracy and state synchronization accuracy but also the extent to which ONDT can provide additional value and the feasibility of deploying ONDT in an existing commercial operational ecosystem.
Such an evaluation will include multi-vendor applicability, dependency on vendor-specific information and additional monitoring, functional integration with existing functions, migration in stages and associated cost, and operational flexibility by open models/interfaces.
\subsection{Emerging Optical Network Services and Ecosystems}
\label{Emerging Optical Network Services and Ecosystems}
Emerging optical network services generally require not only high capacity but also greater flexibility and faster service provisioning.
Examples include DCX, Alien Access Links (AALs), Optical Spectrum as a Service (OSaaS), and multi-domain networks (see~\ref{Ecosystem Architecture}).
These emerging services may be configured with E2E optical connections that combine TRxs and network resources owned by different management entities such as users, network operators, and service providers.
In such a case, network infrastructure, physical-layer information, and control authority will be distributed among multiple authorities, which means that a single management entity may not have a complete understanding of the network state on an E2E basis.
%
In this context, ONDT must capture the required network state through monitoring, channel probes, and DLM while addressing interoperability, information sharing, and responsibility boundaries across vendors and domains.
Thus, ONDT can augment existing operations as well as support new optical services across a variety of management entities and network resources.
\begin{table*}[t!]
\centering
\caption{
Representative Experimental and Field Validations of ONDT for Emerging Optical Network Services and Ecosystems}
\label{tab:ondt_validation_emerging}
\renewcommand{\arraystretch}{1.15}
\setlength{\tabcolsep}{3.2pt}
\scriptsize
\begin{tabularx}{\textwidth}{|
    >{\raggedright\arraybackslash}p{1.20cm}|
    >{\raggedright\arraybackslash}p{1.55cm}|
    >{\raggedright\arraybackslash}p{2.05cm}|
    >{\raggedright\arraybackslash}p{1.95cm}|
    >{\raggedright\arraybackslash}p{3.20cm}|
    >{\raggedright\arraybackslash}p{1.05cm}|
    >{\raggedright\arraybackslash}X|}
\hline
\textbf{Study} &
\textbf{Validation environment} &
\textbf{Main ONDT function validated} &
\textbf{Measurements / observability} &
\textbf{Key quantitative result} &
\textbf{FDD perspective} &
\textbf{Deployment implication / limitation} \\
\hline

OSaaS Probing~\cite{Kaeval2025OSaaS} &
Disaggregated / multi-domain optical service &
Probe-based QoT assessment and service-margin estimation &
QoT probe channels; GSNR at service/domain boundaries &
Probe-based GSNR estimation experimentally validated; quantitative accuracy discussed in the preceding subsection &
\textbf{F/Dp} &
Enables service-feasibility assessment without complete disclosure of internal physical-layer information. \\

\hline

DCX Fast Provisioning~\cite{mano2023first, nishizawa2024fast} &
Field-fiber DCX with heterogeneous TRxs and AALs &
Minimal-probing E2E QoT estimation and mode selection &
Link-by-link BER/GSNR probing and TRx characteristics &
Optical path provisioned in 351~s; estimated--measured Q-factor difference $<0.7$~dB &
\textbf{F/Dy/Dp} &
Enables rapid provisioning across heterogeneous user/carrier equipment; open interfaces and containerized device software improve portability. \\

\hline

DT-Assisted Provisioning~\cite{borraccini2023partiallyDT} &
Multi-vendor, partially disaggregated testbed &
physical-layer-DT-assisted lightpath deployment and recovery &
Device telemetry, GNPy physical-layer-DT, and ONOS &
400-Gb/s request provisioned using 2$\times$DP-16QAM; failure-triggered recovery demonstrated using alternative DP-QPSK paths &
\textbf{F/Dy/Dp} &
Demonstrates vendor-agnostic integration of physical-layer intelligence with open SDN control. \\

\hline

Semi-automatic provisioning~\cite{Nishizawa2024semi-automatic} &
Field-deployed DCI / line-system environment &
DLM/OLS-calibration-based physical characterization and line-system provisioning &
DLM, OLS calibration, and coherent-TRx QoT measurements &
Parameter retrieval, optimization, and provisioning completed within 1~h; QoT error reduced from $\pm0.6$ to $\pm0.3$~dB &
\textbf{F/Dy/Dp} &
Reduces on-site instrumentation and work, and supports remote provisioning of insufficiently characterized infrastructure. \\

\hline

OSNR / GSNR prediction
in brownfield~\cite{Raj2026DLMBrownfield} &
Live 177.5-km multi-vendor brownfield link; 303 service cases &
Measurement-anchored hybrid physics/ML OSNR/GSNR prediction &
DLM sweeps, node power readings, and component models &
Power MAE $\leq0.20$~dB; OSNR MAE $\leq0.39$~dB; GSNR MAE $\leq0.43$~dB &
\textbf{F/Dp} &
Combines reusable physics with ML where device information is unavailable, supporting brownfield and disaggregated services. \\

\hline

DLM multi-domain~\cite{Sasai2025ONTMultiDomain} &
Live production multi-domain network &
E2E state visualization and bottleneck localization across administrative domains &
Commercial 800G TRx only; receiver-based longitudinal monitoring &
First live multi-domain demonstration; power visualization used for bottleneck localization, optimization, and provisioning &
\textbf{F/Dp} &
Provides E2E observability without requiring operators to expose detailed internal-domain information. \\

\hline

\multicolumn{7}{l}{
\textit{F: Fidelity; Dy: Dynamicity; Dp: Deployability. MAE: Mean Absolute Error.}
}
\end{tabularx}
\end{table*}
\subsubsection{Validation for Emerging Services and Ecosystems}
In emerging optical network services, new service models are being studied that will make flexible use of network resources owned by different users, network operators, and service providers and provide optical connectivity and transmission resources as needed (see \ref{Ecosystem Architecture}).
A feature common to these service models is that network infrastructure, TRxs, physical-layer information, and control authority are not necessarily consolidated under a single entity.
As a result, the ability to grasp E2E physical-layer information in a centralized manner cannot be assumed, and it will be important to grasp the necessary network state and service feasibility from available measurements and information and to achieve provisioning and operation across heterogeneous resources.

In OSaaS, for example, a dense wavelength-division multiplexing (DWDM) network operator provides spectral resources to end customers while the selection, operation, and upgrading of TRxs are handled on the customer side.
The possibility therefore exists of transparently connecting multiple domains consisting of different operators, DWDM technologies, components, and channel plans, but on the other hand, demarcation, compatibility monitoring, and QoT assessment will become necessary in contrast to a conventional single-vendor network.
For such an OSaaS environment, a method using QoT probe channels was experimentally demonstrated to evaluate the GSNR of individual links and use that information to estimate E2E transmission feasibility and service margin~\cite{Kaeval2021QoT}.
Of importance here is that this technique can evaluate service feasibility based on a common QoT metric obtained from active probing without the sharing of complete physical-layer information.
This is seen as an approach that can enhance service-level interoperability and Deployability in a multi-operator and disaggregated environment.
This measurement-based approach was further extended to DCX, where automatic provisioning was demonstrated in an environment that includes AALs whose physical parameters are not sufficiently understood by the carrier and user TRxs having different characteristics~\cite{mano2023first, nishizawa2024fast}.
In DCX, it is necessary to handle diverse routes, wavelengths and TRx combinations between multiple data centers, cloud systems, and multi-access edge computing (MEC) nodes, so it is not realistic to characterize all candidate paths beforehand.
A method was introduced that selects an appropriate transmission mode by characterizing individual links using QoT probe channels and estimating E2E GSNR from those results without directly measuring operational routes and wavelengths themselves.
%

These demonstrations show that ONDT-related functions can be deployed in actual network environments using open software interfaces.
On the other hand, executing optimization and control based on a physical-layer model on such a software infrastructure requires that the physical state of the target OLS be appropriately modeled and that a digital representation that can be used for network control be constructed.
With this in mind, the integration of a physical-layer digital twin into open and disaggregated network control was experimentally demonstrated for QoT-aware network optimization and control~\cite{borraccini2023partiallyDT}.
However, in expanding such model-based operation to an actual field environment, an important issue in terms of Deployability is how to actually obtain fiber, amplifier, and other physical-layer parameters from the actual network infrastructure for input to GNPy.
Semi-automatic provisioning was demonstrated on a field-deployed OLS by combining DLM-based monitoring with physical-parameter calibration, visualization, and QoT-aware optimization using a physical-layer model~\cite{Nishizawa2024semi-automatic}. 

For practical DCX service deployment, the field trial addressed operational requirements such as minimizing additional monitoring equipment and on-site interventions, enabling remote design and provisioning, and completing line-system provisioning within a limited time.
It validated the process from physical-parameter extraction to line-system optimization and provisioning, including visualization of power profiles and physical-layer characteristics such as QoT dependence on fiber input power, while considering compatibility with existing operational practices.
This measurement-assisted approach was extended to OSNR/GSNR prediction in brownfield optical links by combining DLM with a hybrid physics/ML digital twin~\cite{Raj2026DLMBrownfield}. The approach retained physics-based models for fiber propagation and NLI, applied learned models to device components with uncertain characteristics, and adapted the models using DLM observations.
This model is regarded to be an example of allocating Fidelity in a fit-for-purpose manner.
In other words, it selectively introduces detailed modeling and learning for components that have a large impact on QoT prediction but whose characteristics are uncertain, and it achieves both Fidelity and Deployability by using reusable physics-based models.

Optical network tomography was demonstrated in a multi-domain environment of a live production network to obtain E2E information on an optical link using receiver-based longitudinal monitoring based on commercial coherent TRxs~\cite{Sasai2025ONTMultiDomain}.
This method assumes that such domains will not mutually disclose their detailed configurations and physical parameters.
It therefore obtains a longitudinal optical-power profile from end-point observations while also visualizing the state of transmission across domain boundaries and enabling bottleneck localization.
This demonstration is viewed as extending the role of ONDT from a stage of complementing uncertain device/link parameters within a single operational environment to a stage of establishing E2E observability in an environment in which physical-layer information is dispersed across multiple domains.

The validations described above show that the role of ONDT in emerging optical services is not limited to improving the accuracy and automation of existing network operation.
Rather, by obtaining the network state required for service operation from available measurements and information and constructing its digital representation, and by using that representation across different devices, operators, and administrative domains via interfaces and software architectures, ONDT can become a platform supporting optical services that have been difficult to achieve under past administrative boundaries and information constraints.
However, these results were obtained under specific service scenarios, network environments, and operational conditions and do not establish the general Deployability of ONDT for emerging services.
To facilitate the widespread use of ONDT in commercial networks from here on, it is important to validate the extent to which these operational benefits can be reproduced in diverse network scales, vendors, generations, administrative domains, and operational conditions.
In short, ONDT must be further evaluated in E2E service operation that includes factors such as service continuity, demarcation of responsibility, costs, and integration with existing operations.

\section{Conclusion}
\label{Conclusion}
This survey presented a uniform ONDT paradigm that treats the historical development of the digitalization of the optical network, multi-dimensional architecture, physical-layer modeling and monitoring, network control, data and control governance, as well as AI-agent-assisted operation as a consistent flow. The ONDT is not a single model or software platform but rather a system that constructs and synchronizes a reliable digital representation of the physical network for use in prediction, optimization, and network operation. Its architecture, moreover, is organized into the four complementary design dimensions of Ecosystem, Functional, Control, and Implementation so that a variety of ONDT implementations can be positioned in a common design space.

A main contribution of this survey is a FDD framework.
Here, it is not a matter of maximizing these three requirements in isolation.
In a real-world ONDT, it is important to balance these three requirements depending on the target network and operational objectives.

The experiments and field validations presented in this study also reveal different roles for ONDTs depending on deployment scenarios.
In the case of conventional optical networks, commercial services and operating systems have already been established and running for some time, so an evaluation must be performed from the viewpoint of what kind of additional value ONDT will be able to provide to the existing system.
In the case of emerging optical services and ecosystems, however, ONDT may play a different role.
Field demonstrations have shown that a digital representation required for service operations can be constructed even if complete physical-layer information cannot be obtained beforehand by combining channel probing, physical-layer digital twin, DLM or network tomography, physical-parameter calibration, and open and cloud-native control.
From this perspective, Deployability is not just about deploying ONDT into existing networks, but also about enabling new optical services and operational ecosystems with limited or distributed information availability.

ONDT and AI agents should be seen as having complementary roles rather than one being an alternative to the other.
While ONDT possesses physical consistency and provides a digital representation synchronized with the physical network, AI agents use that representation to perform reasoning and planning, use tools, and make decisions.
Maintaining a clear division of roles in this way is important from the viewpoints of transparency, traceability, safety, and operational governance as network control becomes increasingly autonomous.
As described above, the practical value of ONDT does not simply lie in the construction of a digital replica that is as detailed as possible. Rather, it lies in the ability to provide knowledge of the physical network that is sufficient and reliable under the accuracy, time scale, and deployment cost required for the target operation.
The FDD framework and the systematic organization of architecture, enabling technologies, and validation presented in this survey can serve as a common platform for the design and evaluation of future ONDT systems.


%

\section*{Acknowledgment}
All technical knowledge, ideas, and figures were created by human authors. The authors used OpenAI's ChatGPT and Anthropic Claude for information search, language and structural editing, and text compression throughout the paper. The work of Fraunhofer HHI has been partially funded by the BMFTR project SHINKA (FKZ 16KIS2365). This work was supported in part by NSF under grants 2112562, 2330333, 2443137, 2450567, 2503010, and 2549359.

\ifCLASSOPTIONcaptionsoff
  \newpage
\fi



%

\bibliographystyle{IEEEtran}
\bibliography{IEEEabrv,ref_compact_visualization}

%

\raggedbottom

\begin{IEEEbiographynophoto}{Hideki Nishizawa}
is a Senior Research Engineer at NTT Network Innovation Laboratories, Japan. He received B.S. and M.S. degrees from Chiba University and a Ph.D. from Hokkaido University.
\end{IEEEbiographynophoto}
%
\begin{IEEEbiographynophoto}{Enrico Miotto}
is a Ph.D. student with the PLANET Lab at Politecnico di Torino, Italy. He received B.Sc. cum laude from University of Padua and M.Sc. cum laude from Politecnico di Torino.
\end{IEEEbiographynophoto}
%
\begin{IEEEbiographynophoto}{Andrea Rosso}
is a Ph.D. student with the PLANET Lab at Politecnico di Torino, Italy, where he previously earned his B.Sc. and subsequently his M.Sc. cum laude.
\end{IEEEbiographynophoto}

\begin{IEEEbiographynophoto}{Takahito Tanimura}
is a Manager at the Edge Intelligence Research Department, R\&D Group, Hitachi, Ltd., Japan. He received B.S. and M.S. degrees from Tokyo Institute of Technology (now Institute of Science Tokyo), and a Ph.D. from the University of Tokyo. 
\end{IEEEbiographynophoto}
%
\begin{IEEEbiographynophoto}{Zehao Wang}~is currently pursuing his Ph.D. degree in the Department of Electrical and Computer Engineering at Duke University.
\end{IEEEbiographynophoto}
%
\begin{IEEEbiographynophoto}{Mingzhe Han}~is a Ph.D. student in the Department of Electrical and Computer Engineering at Duke University. 
\end{IEEEbiographynophoto}

\begin{IEEEbiographynophoto}{Hiroyuki Ishihara}~is a research engineer at NTT Network Innovation Laboratories, Japan. He received the B.E. degree from National Institute of Technology, Matsue College and the M.S. degree from Kyoto University.
\end{IEEEbiographynophoto}

\begin{IEEEbiographynophoto}{Andrea D'Amico}\;
is a Researcher in the ONS Department at NEC Laboratories America. He holds a Ph.D. in Telecommunications Engineering from Politecnico di Torino. 
\end{IEEEbiographynophoto}

\begin{IEEEbiographynophoto}{Agastya Raj}\;
is a PostDoctoral Researcher in Trinity College Dublin. His research focuses on data-driven modelling and digital twins of optical networks, ML for optical network automation, and fibre-optic sensing.
\end{IEEEbiographynophoto}

\begin{IEEEbiographynophoto}{Rocco D'Ingillo} received the M.Sc. degree in Electronic Engineering (2021) and the Ph.D. degree in Teleommunications Engineering (2025) from Politecnico di Torino.
\end{IEEEbiographynophoto}

\begin{IEEEbiographynophoto}{Takeo Sasai}
  received the B.E., M.E. and Ph.D. degrees from the University of Tokyo, Japan, in 2015, 2017, and 2025 respectively. In 2017, he joined NTT Laboratories, Yokosuka, Japan.
\end{IEEEbiographynophoto}
%
\begin{IEEEbiographynophoto}{Giacomo Borraccini}\;
currently holds the position of research fellow at Trinity College Dublin. 
His research focuses on optical networks at the physical layer, mainly on optical transmission at the system level in terms of modeling and control.
\end{IEEEbiographynophoto}
%
\begin{IEEEbiographynophoto}{Toru Mano}
is a research engineer at NTT Network Innovation Laboratories. He received his B.E. and M.E. degrees from the University of Tokyo and a Ph.D. from Hokkaido University.
\end{IEEEbiographynophoto}
%
\begin{IEEEbiographynophoto}{Renato Ambrosone}\;
is a Ph.D. student in the PLANET group at Politecnico di Torino. He earned an M.Sc. in Computer Engineering from Politecnico di Torino.
\end{IEEEbiographynophoto}
%
\begin{IEEEbiographynophoto}{Behnam Shariati}\;
is a the Deputy Head of Data Analytics and Signal Processing Group at Fraunhofer HHI. He leads the AI for Photonics topical area that develops state-of-the-art solutions for network automation.
\end{IEEEbiographynophoto}
%
\begin{IEEEbiographynophoto}{Oscar González-de-Dios}
is the Head of SDN deployments for Transport Networks at Telefonica Global CTIO. He received the M.S. and Ph.D. degrees from the University of Valladolid.
\end{IEEEbiographynophoto}
\begin{IEEEbiographynophoto}{Arturo Mayoral López de Lerma} received the Ph.D. degree in Telecommunications Engineering (2019) from Universitat Politecnica de Catalunya.
\end{IEEEbiographynophoto}

\begin{IEEEbiographynophoto}{Jun-ichi Kani}
is a Senior Distinguished Researcher at NTT Access Network Service Systems Laboratories, Japan. He received the B.E., M.E., and Ph.D. degrees from Waseda University.
\end{IEEEbiographynophoto}
\begin{IEEEbiographynophoto}{Manabu Yoshimoto}
is a Senior Manager at he NTT R\&D Center, focusing on optical network operations and AI-driven automation. He received his bachelor's degree in mechanical engineering from Tohoku University.
\end{IEEEbiographynophoto}

\begin{IEEEbiographynophoto}{Tingjun Chen}
is the Nortel Networks Assistant Professor of Electrical and Computer Engineering at Duke University. He received his B.Eng. degree from Tsinghua University and Ph.D. degree from Columbia University. His research focuses on next-generation wireless, mobile, and optical networked systems.
\end{IEEEbiographynophoto}

\begin{IEEEbiographynophoto}{Marco Ruffini}
is full professor at Trinity College Dublin, leading the OpenIreland lab and the Optical and Wireless networks research group.
\end{IEEEbiographynophoto}
%
\begin{IEEEbiographynophoto}{Daniel Kilper}
is Professor of Future Communication Networks and Director of the CONNECT Centre at Trinity College Dublin, Ireland. He received M.S. and Ph.D. degrees in physics from the University of Michigan.
\end{IEEEbiographynophoto}


\begin{IEEEbiographynophoto}{Vittorio Curri}
is Full Professor at Politecnico di Torino, Italy, where he leads the Open PLANET Lab. He received his Laurea and Ph.D. degrees from Politecnico di Torino. He is an IEEE, OPTICA, and AAIA Fellow.
\end{IEEEbiographynophoto}





\end{document}